\documentclass[11pt, english]{article}
\usepackage[utf8]{inputenc}
\usepackage{amssymb,amsmath}
\usepackage{color}
\usepackage{soul}
\usepackage{graphicx}
\usepackage{gensymb}
\usepackage{textcomp}
\usepackage{subfig}
\usepackage{algorithm2e}
\usepackage{authblk}
\usepackage{multirow}
\usepackage{array}
\usepackage{mathtools}
\usepackage{verbatim}
\definecolor{Red}{rgb}{1,0,0}
\definecolor{Green}{rgb}{0,1,0}
\definecolor{Blue}{rgb}{0,0,1}
\definecolor{Orange}{rgb}{1,0.5,0}

\title{Towards Real-Time Estimation of Solar Generation From Micro-Synchrophasor Measurements}

\author[1]{Emre C. Kara\thanks{emrecan@slac.stanford.edu}}
\author[2]{Ciaran M. Roberts}
\author[3]{Michaelangelo Tabone}
\author[4]{Lilliana Alvarez}
\author[3]{Duncan S. Callaway}
\author[2]{Emma M. Stewart}
\affil[1]{Grid Integration Systems and Mobility Group, SLAC National Accelerator Laboratory, Menlo Park, CA, USA}
\affil[2]{Energy Storage and Distributed Resources Division, Lawrence Berkeley National Laboratory, Berkeley, CA, USA}
\affil[3]{Energy and Resources Group, University of California, Berkeley, CA, USA}
\affil[4]{Riverside Public Utilities, Riverside, CA, USA}

\begin{document}

\maketitle

\begin{abstract}
This paper presents a set of methods for estimating the renewable energy generation downstream of a measurement device using real-world measurements. First, we present a generation disaggregation scheme where the only information available for estimation is the micro-synchrophasor measurements obtained at the substation or feeder head. We then propose two strategies in which we use measurements from the substation as well as a proxy solar irradiance measurement. Using these two measurement points, we first propose a multiple linear regression strategy, in which we estimate a relationship between the measured reactive power and the load active power consumption, which are then used in disaggregation. Finally, we expand this strategy to strategically manage the reconstruction errors in the estimators. We simultaneously disaggregate the solar generation and load. We show that it is possible to disaggragate the generation of a 7.5 megawatt photovoltaic site with a root-mean-squared error of $\approx$ 450 kilowatts.
\end{abstract}

\section{Introduction}
\label{sec:introduction}
The distribution grid's role in power supply is undergoing a paradigm shift with the influx of distributed energy resources (DER). Historically, the distribution grid transferred power from the transmission network to the customer. This transfer of power was primarily uni-directional. This, however, is set to change as the grid tends from a centrally controlled system to one more distributed in nature. As DER further penetrates the grid, the complexity of real-time operation of the power system rises. Large, centralized fossil-fuel plants are being replaced by smaller, geographically dispersed, intermittent resources such as wind and solar. 

Visibility into the instantaneous generation of these intermittent resources is critical to maintain real-time balance between power generation and load. In this study, we seek to gain this visibility by leveraging the instantaneous active and reactive power demand of a feeder, alongside an irradiance proxy measurement, to estimate the overall photovoltaic (PV) generation at the feeder. The active and reactive power demand are determined from current and voltage phasor measurements from micro-synchrophasor measurement units ($\mu$PMUs). These devices, recently developed by Power Standards Lab (PSL) \cite{upmu_site}, are deployed at Riverside Public Utility (RPU) with a sampling frequency of 120 Hz. They have the potential to enable advanced data analytics, which facilitate the transformation of the distribution grid into a more dynamical and bidirectional---yet still reliable---system \cite{scoping_study}. Researchers have proposed using synchronized phasor measurements to estimate instantaneous generation and load \cite{von2014micro} \cite{sexauer2013phasor}. This approach, however, has yet to be explored. To the best of our knowledge, this is the first work that seeks to utilize $\mu$PMU measurements to estimate the instantaneous PV generation downstream from the measurement device. 

In an actively managed distribution grid, operators need to be able to access detailed status information about the system at any given moment. Typically, system operators gather this knowledge via state estimation. The benefit of phasor measurement units (PMUs) in conducting distribution system state estimation has been explored in the literature \cite{liu2012trade, macii2015impact, sarri2012state}, with \cite{macii2015impact} specifically focusing on their benefit in the presence of a large PV penetration. The authors noted that the placement of just a single PMU reduced the average root-mean-squared-error (RMSE) by a factor ranging from 2 to 7, dependent on the total vector error of the measurement device and level of PV penetration. In this work, we seek to further expand the use of PMUs in estimating the instantaneous solar generation at the feeder, which in turn could improve the accuracy of state estimation. 

Estimating instantaneous generation also allows visibility into the \emph{masked load} of a feeder. Understanding this masked load allows the system operator to anticipate steep ramp rates---e.g., as the PV generation ramps down in the evening and the system approaches the residential peak---as well as procure the necessary resources to protect supply reliability should one or more distributed generation (DG) units trip. 

Our main purpose in this study is to investigate ways to disaggregate solar generation from aggregate information obtained at the substation level. In particular, we are interested in disaggregation schemes that require the least amount of information and can provide accurate, close to real-time monitoring of the actual PV generation on the feeder.
For this purpose, we first propose a disaggregation scheme where the only information available is the $\mu$PMU measurements obtained from the substation. In this strategy, we leverage heuristics on PV inverter and load characteristics to estimate the total amount of PV generation in the feeder. We refer to this methodology as \emph{power factor-based estimator} (PFBE). 

We then propose two strategies in which we assume that measurements from the substation and a proxy solar irradiance measurement are available. In this case, the proxy solar irradiance corresponds to the active power output of a nearby solar installation on a separate circuit. Using these two measurement points, we first propose a multiple linear regression strategy, in which we estimate the relationship between the active and reactive power consumption of the load, which then is used in disaggregation. We refer to this estimator as the \emph{linear estimator} (LE). We achieve an RMSE of 6\% of installed capacity across all sky conditions by making assumptions regarding the attribution of model errors.
Finally, we expand this strategy to a contextually supervised source separation, building on the methodology proposed in~\cite{wytock2013contextually}. In this second strategy, we simultaneously disaggregate the PV generation and load, and estimate an effective power factor. This allows for the previous assumptions on the errors to be removed and a more systematic attribution of model errors. This revised methodology allows for further improvement on the accuracy given a more representative irradiance proxy. We refer to this methodology as the \emph{contextually supervised generation estimator} (CSGE). 

In comparison to the approach presented here, some earlier work has focused on using real-time irradiance measurements and known installed PV capacity to estimate PV generation. This model-based approach attempts to leverage one or more irradiance sensors and model effects such as cell temperature to estimate the aggregate generation of a cluster of PV sites. In Section~\ref{sec:rel_work}, we demonstrate that such irradiance measurements lead to an inherent accuracy barrier that makes this approach unsuitable for producing accurate, real-time estimations. 

An alternative estimation approach exploits the requirement of metering for PV sites above a specified capacity threshold. When present, however, these meters typically offer low-granularity measurements and are often not required on residential installments, thus offering limited visibility. Although there is increasing interest in real-time monitoring and control of solar inverters by utilities and distribution grid operators, it requires a higher bandwidth communications infrastructure, which requires greater implementation costs~\cite{reiter2015industry}.
The methodology we propose possesses the potential to overcome the granularity and cost barriers to real-time solar generation monitoring.

This paper evaluates three methods for PV disaggregation in progression; where each method relies less on assumptions of PV or load power factor and more on other exogenous observations. 
No method requires information on the amount installed PV capacity or plane of array geometry. Specifically, our main contributions are:
(i) an exploration of three methodologies for PV generation disaggregation, (ii) a systematic analysis of PV generation estimation errors in these methodologies, (iii) a PV estimation strategy (CSGE), which achieves an RMSE of 6\% of installed capacity across all sky conditions, allows load and PV power factors to vary based on observations, and hence, does not rely on the reactive power behavior of individual inverters.

In the following sections, we first introduce the related work and the dataset, and then explore the strategies briefly described above.


\section{Related Work} \label{sec:rel_work}

A common method for predicting the output of PV systems is a model-based approach \cite{gotseff2014accurate, alonso2015beam}, which uses a model, composed of the design characteristics of the PV array and the irradiance measurements, to estimate PV generation. In order to understand this approach, and how the proposed methodology in this paper offers a significant improvement for the purpose of real-time estimation of PV generation, we briefly discuss its limiting factors. The critical input to these models is the plane-of-array (POA) irradiance. The POA irradiance on a tilted plane, $E_{s}$, whose tilt is $s$ degrees from the horizontal plane can be calculated using Equation~\eqref{poa_irr}~\cite{gueymard2009direct}

\begin{equation}
E_{s} = E_{dni} cos \theta + E_{dhi} R_{d} + \rho E_{ghi} R_{r}
\label{poa_irr}
\end{equation}

\noindent where $E_{dni}$ is the direct normal irradiance (DNI), $\theta$ is the angle of incidence of the sun rays on the array, $E_{dhi}$ is the diffuse horizontal irradiance (DHI), $R_{d}$ is the diffuse transposition factor, $\rho$ is the foregrounds albedo, $E_{ghi}$ is the global horizontal irradiance (GHI), and $R_{r}$ is the transposition factor for ground reflection. Given site-specific measurements of both DNI and DHI, the authors of~\cite{gueymard2009direct} estimate the $E_{s}$ with approximately 5\% accuracy for a 40$^{o}$ tilted south-facing panel. The issue, however, is that accurate measurements of both DNI and DHI are costly to obtain and therefore, typically, only the global horizontal irradiance, E$_{ghi}$, is measured \cite{gueymard2009direct} \cite{gueymard2010progress}. In this case both the DNI and DHI need to be estimated from a given GHI measurement. A total of ten models were examined in~\cite{gueymard2009direct} for estimating both the DNI and DHI, given site-specific GHI measurements, with typical $E_{s}$ RMSE of $\approx$ 10\% for all sky conditions for a 40$^{o}$ tilted south-facing panel and an RMSE of $\approx$ 13\% for a two-axis tracking plane. Furthermore, in a follow-up study, \cite{gueymard2010progress}, focused on the estimation of DNI given site-specific GHI, the author concluded that no model performs consistently well over all-sky conditions and that, generally, the more recent models did not seem to outperform models introduced 30 years ago. The use of satellite-derived estimations appears to offer no significant improvement in performance~\cite{alonso2015beam} \cite{jamaly2013aggregate}. The inaccuracy in estimating the DNI, and lack of site-specific GHI measurments,  render real-time irradiance measurements suboptimal for real-time disaggregation of DG. In this work, we use an irradiance proxy, in the form of active power output of a single PV installation, to overcome this issue by negating the need to explicitly estimate each of the inputs to Equation~\eqref{poa_irr}. 

Jamaly et al.~\cite{jamaly2013aggregate} develop an approach to anticipate aggregate PV ramp rates for 86 DG systems utilizing satellite irradiance data and demonstrate an RMSE of $\approx$ 20\%. The authors conclude that the primary source of this error stemmed from the inaccuracy of the satellite-estimated irradiance rather than the model employed to estimate power production. This is in line with the conclusions of \cite{gueymard2009direct} and \cite{gueymard2010progress} regarding POA irradiance estimation inaccuracy. This inaccuracy, coupled with the low frequency of the data, 30-minute resolution for this particular study, leaves significant room for improvement for the purpose of real-time disaggregation. 

In addition to uncertainties in the estimation of the POA irradiance, there are further sources of potential error in the model-based approach, which may be difficult to account for accurately. One such source of error is the estimation of the soiling losses, arising from the accumulation of particulate matter on PV systems. These soiling losses averaged 0.051\%/day during dry periods for a sample of 1286 residential and commercial PV sites in California \cite{mejia2013soiling}. There were significant differences, however, between rates with 26\% of sites having losses above 0.1\%/day while systems with a tilt angle $<$ 5$^{o}$ had mean soiling losses of 0.18\%/day. There were no statistically significant differences observed by region, making such a phenomenon more difficult to model. The effect of ground conditions, and similar particulate matter contributors, coupled with cleaning methods, when employed, introduce difficulties in estimating the soiling losses of aggregate PV sites such as a residential feeder. Issues such as the degradation of PV arrays and effect of cell temperature further impact the reliability of feeder-specific generation estimations. 

In this paper, we propose a methodology that captures the aggregate effect of these factors, rather than explicitly modelling each one individually, thus improving the estimation accuracy. This proposed methodology does not rely on the inference of POA irradiance based on GHI measurements, and outperforms model-based approaches, particularly for the case of sparsely distributed GHI sensors.


\section{Data}
\begin{figure}
\centering
\includegraphics[width=3.5in]{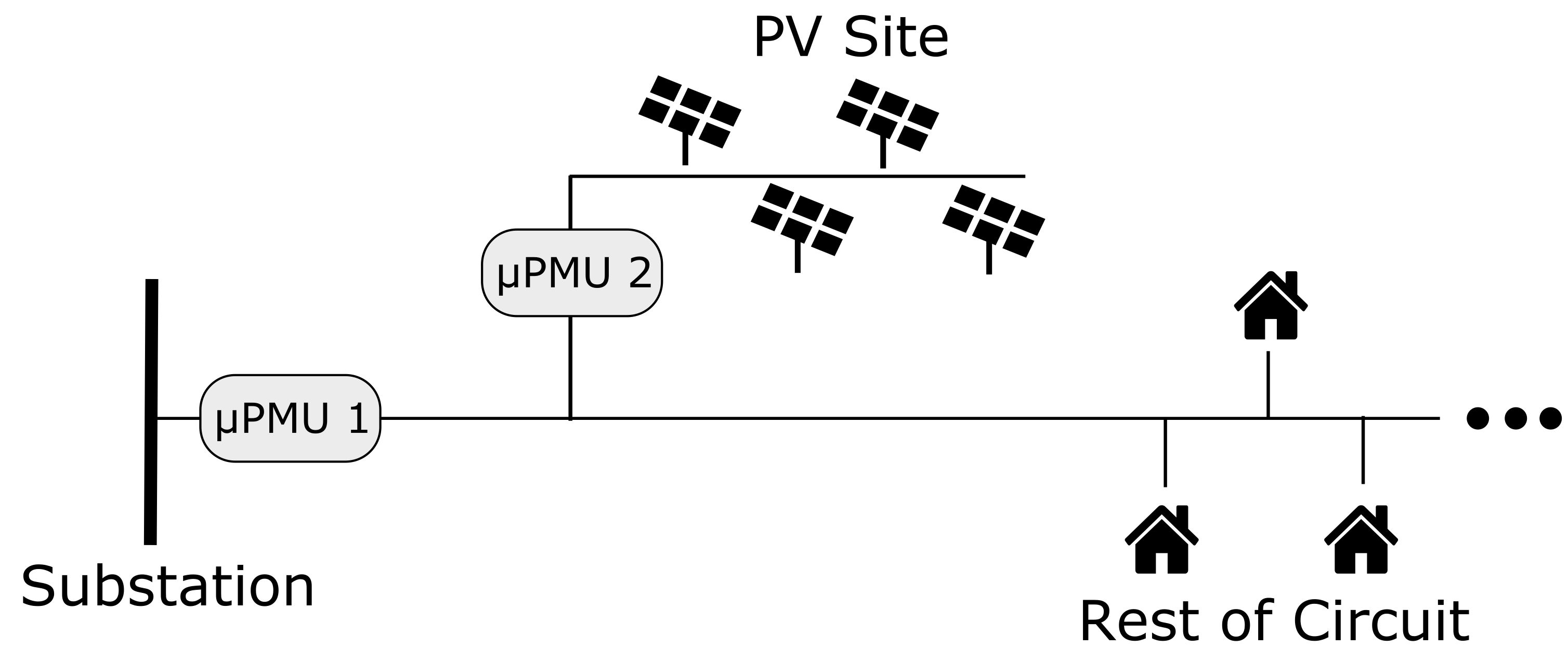}
\caption{Simplified circuit diagram showing measurement locations}
\label{fig:RiversideMap}
\end{figure}

\begin{figure}
\centering
\subfloat[Active power calculated from $\mu$PMU$_1$ measurements\label{upmu1realpower}]{%
  \includegraphics[width=1\textwidth]{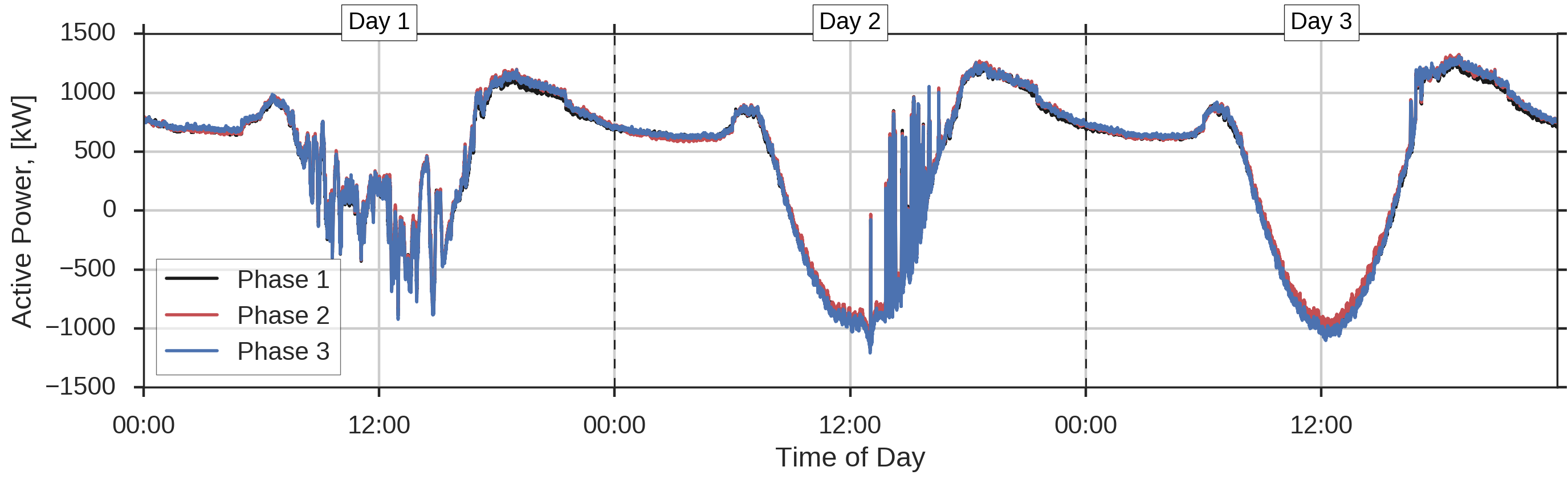}
}    
\hfill
\subfloat[Reactive power calculated from $\mu$PMU$_1$ measurements\label{upmu1reacpower}]{%
  \includegraphics[width=1\textwidth]{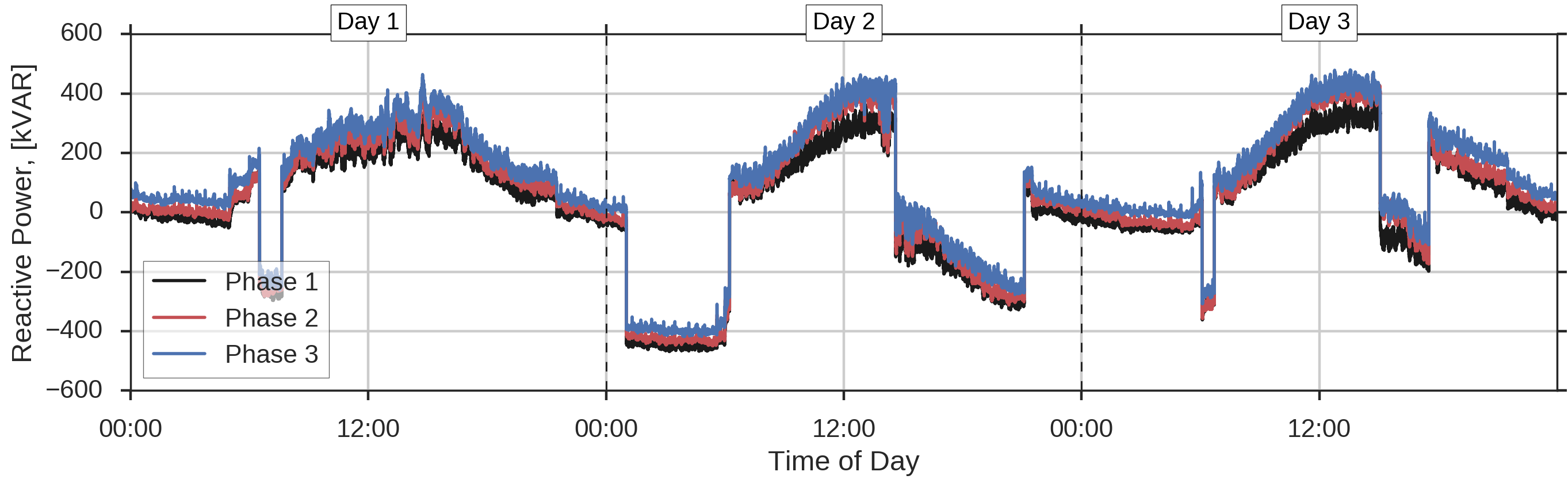}
}
\hfill
\subfloat[Active power calculated from $\mu$PMU$_2$ measurements\label{upmu2realpower}]{%
  \includegraphics[width=1\textwidth]{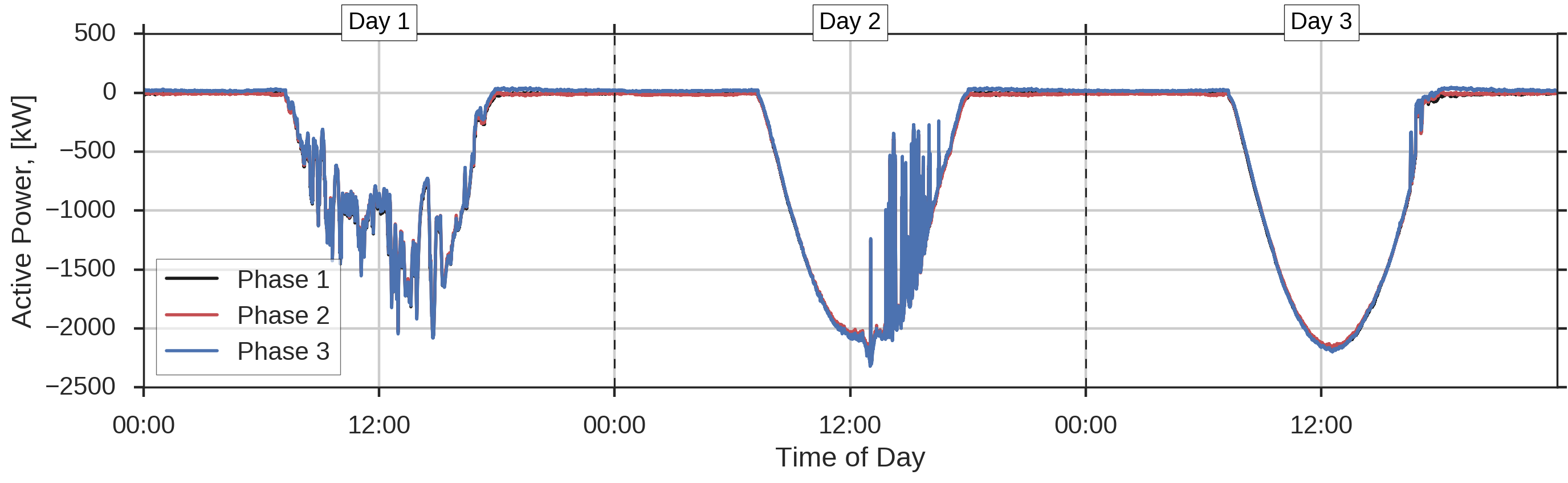}
}
\hfill
\subfloat[Reactive power calculated from $\mu$PMU$_2$ measurements\label{upmu2reacpower}]{%
  \includegraphics[width=1\textwidth]{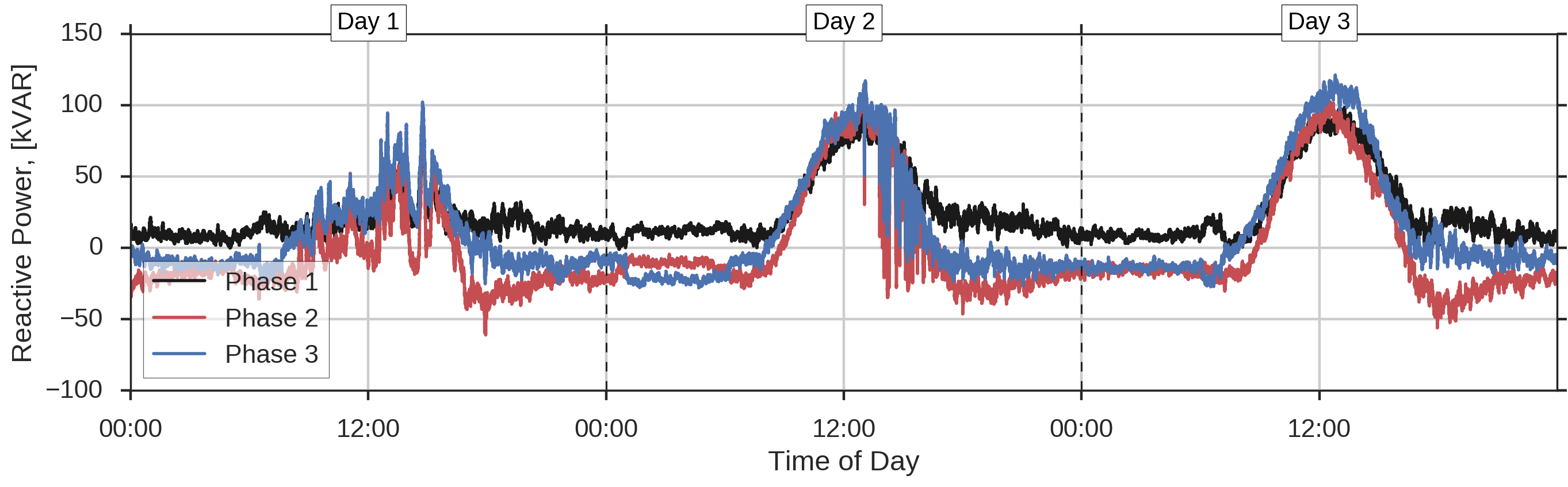}
}
\caption{Active and reactive power calculated for all phases using $\mu$PMU measurements}
\label{fig:power}
\end{figure}

The main dataset used in this study is obtained from two $\mu$PMUs located in Riverside Public Utility (RPU) territory. The first $\mu$PMU ($\mu$PMU$_1$) is located at the substation, while the second $\mu$PMU ($\mu$PMU$_2$) is located at the point of interconnection of a 7.5 megawatt capacity PV generation site further downstream from the substation. It is important to note that this PV generation site is the only generation asset located at this substation. Figure~\ref{fig:RiversideMap} shows the location of both $\mu$PMUs in the distribution grid. Each $\mu$PMU outputs the current and voltage phasors for each phase, from which we obtain active and reactive power, at a rate of 120 Hz. Figure~\ref{fig:power} shows active and reactive power values---calculated using measurements obtained from $\mu$PMU$_1$ and $\mu$PMU$_2$.

In this paper, we use the measurements obtained from $\mu$PMU$_2$ for validation purposes only. As discussed in Section~\ref{sec:introduction}, we use measurements from $\mu$PMU$_1$ for the PFBE to estimate the aggregate PV generation at the substation. For the LE and the CSGE---in addition to measurements from $\mu$PMU$_1$---we use an irradiance proxy measurement to improve our estimates obtained using the PFBE. Specifically, we use active power generation measurements of a nearby PV system at the University of California Riverside's Center for Environmental Research and Technology (CE-CERT) microgrid~\cite{cert_setup} as proxy measurements in this paper. We refer to this measurement as the \emph{irradiance proxy} in the rest of this paper. The PV system that provides the irradiance proxy measurements is located approximately four miles from the monitored substation. The PV system at the CE-CERT microgrid is not connected to the same feeder that we monitor with $\mu$PMUs, thus its generation does not affect net-load at the substation. 

\begin{figure}
    \centering
    \includegraphics[width=1\textwidth]{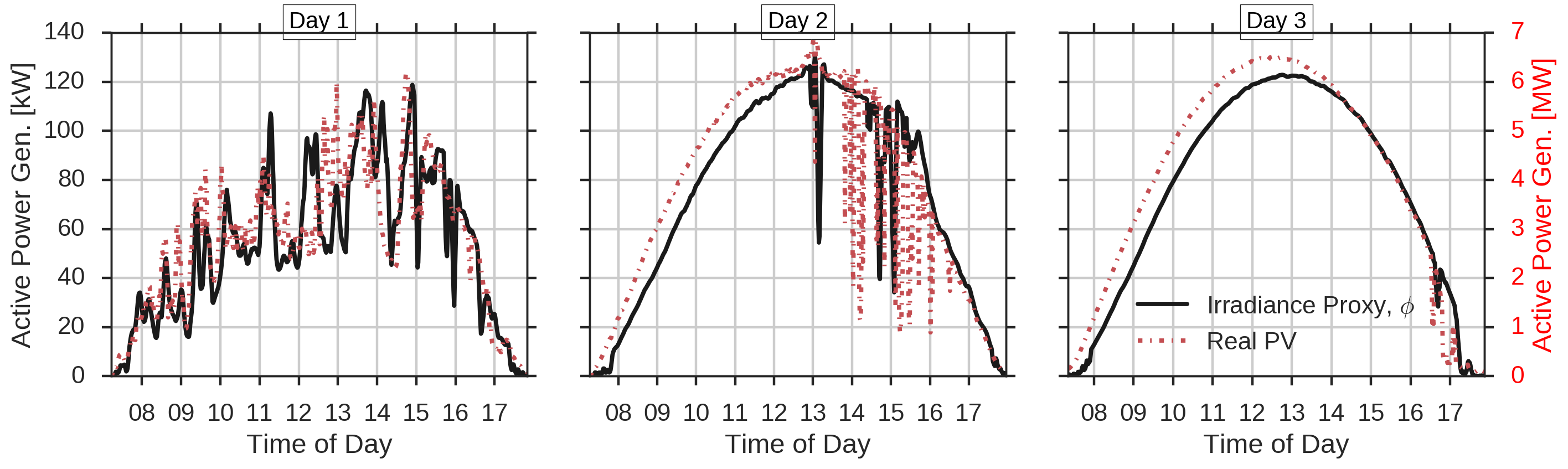}
    \caption{Proxy irradiance measurements obtained from a nearby PV site and the PV generation active power calculated from $\mu$PMU$_2$ measurements.} 
    \label{fig:IrradianceProxy}
\end{figure}

For generation estimation purposes, first, we time-synchronize all measurements used in this study. Specifically, we down-sample active and reactive power readings obtained from $\mu$PMUs to once every minute using interval averages. The motivation for this is two-fold: (i) as demonstrated in \cite{haaren2014empirical}, it is the minute scale which is of primary interest when studying the variability of geographically dispersed or large individual PV sites, and (ii) once every minute sampling rate is the lowest common sampling rate for all the data streams used in this study. Figure~\ref{fig:IrradianceProxy} shows the PV generation at the CE-CERT microgrid and the PV generation calculated from $\mu$PMU$_2$ phasor measurements on separate axes for the same time period. It is possible to see the differences in cloud cover between the two measurement points. In Section~\ref{sec:formulation}, we further discuss the effect of time resolution on our method's ability to recover true solar generation. Furthermore, many previous studies show that high-frequency variations in PV generation are less spatially correlated than lower-frequency variations \cite{mills_implications_2010,perez_short-term_2012,tabone_parameterizing_2013}. To prevent high frequency variations from lessening the predictive power of the nearby PV system, we remove them using a five-minute moving average filter.

\section{Power Factor Based Estimator}
Neglecting losses in the network, for the generic case of a $\mu$PMU measuring the power demand of a feeder, the measured apparent power, $S^{i}_{PMU,t}$, can be expressed as a summation of the apparent power of the load,$S^{i}_{Load,t}$, and the apparent power of installed PV, $S^{i}_{PV,t}$. Equation~\eqref{eq:ComplexPower} represents this for each phase $i$ at time $t$:
\begin{equation}
    S^{i}_{PMU_1,t}=S^{i}_{Load,t}+S^{i}_{PV,t}
\label{eq:ComplexPower}
\end{equation}
Expanding on Equation~\eqref{eq:ComplexPower}, we formulate Equation~\eqref{eq:PowerEquality}:
\begin{equation}
    \overbrace{
    \begin{bmatrix}
    cos{\Phi^{i}_{Load,t}} & cos{\Phi^{i}_{PV,t}}\\
    sin{\Phi^{i}_{Load,t}} & sin{\Phi^{i}_{PV,t}}
    \end {bmatrix}}^{A}
    \times 
     \begin{bmatrix} 
    |S^{i}_{Load,t}|\\ 
    |S^{i}_{PV,t}|    
     \end{bmatrix}
    =  
    \begin{bmatrix}
    P^{i}_{PMU,t}\\ 
    Q^{i}_{PMU,t} 
    \end{bmatrix}
\label{eq:PowerEquality}
\end{equation}
where $P^{i}$ and $Q^{i}$ correspond to the measured active and reactive power on phase $i$, respectively, and $\Phi^{i}$ denotes the phase difference between the voltage and current, whereby cos($\Phi^{i}$) is the power factor.

In Equation~\eqref{eq:PowerEquality}, the measured phase angle terms, $\Phi^{i}_{PMU,t}$ and the apparent power term, $S^{i}_{PMU,t}$ can be obtained in real time from $\mu$PMU$_1$ measurements. Hence, the right-hand side of Equation~\eqref{eq:PowerEquality} is available to the PFBE. The goal is to obtain the unobserved load and PV apparent power values, $S^{i}_{Load,t}$ and $S^{i}_{PV,t}$, respectively. For that, we must know the phase angle differences, $\Phi$, and in order to disaggregate the unobserved apparent power terms, this matrix must be invertible. Furthermore, we make the following assumptions on load and PV generation behavior for the PFBE. 
First, we assume that the load and PV generation power factors are constant. Since a large number of loads are fed from this distribution feeder, we expect the fluctuations in load power factor to be minimal, hence we assume it is constant throughout the estimation period. Similarly, since the PV generation is connected through a controllable inverter, we assume that the inverter is set to provide active power only (i.e., $\Phi^{i}_{PV,t}=180^\degree$). This assumption is based on {current policies}, whereby DG is not required to provide voltage support, and therefore will seek to maximize its revenue by maximizing its active power output. Then, to make the matrix $A$ invertible, it is clear that cos$\Phi^{i}_{PV,t}$sin$\Phi^{i}_{Load,t}$ must be non-zero. Hence, sin$\Phi^{i}_{Load,t}$ must be non-zero. Therefore, the matrix $A$ is invertible for $\Phi^{i}_{Load,t}\neq\{90^\degree,270^\degree\}$.

\paragraph{Estimating Load Power Factors}
The power factor of load, cos$\Phi_{Load,t}$, can be estimated in various ways. For the PFBE, we propose to learn a representative power factor value from the $\mu$PMU$_1$ measurements obtained during periods where there is no PV generation in the system (i.e. overnight). Thus, we inherently assume that the overnight power factor values are also representative of the load power factor throughout the day. To obtain a representative value for the load power factor, we first need to account for any switching devices, such as capacitor banks, that might affect the observed power factor values. As shown in Figure~\ref{upmu1reacpower}, there are five distinct capacitor bank switching events. In the following section, we introduce a threshold-based strategy to detect and account for the capacitor bank switching impacts on reactive power to estimate the load power factor. 

\subsection{Detecting and Compensating for Capacitor Bank Switching}
\begin{figure}[h]
    \centering
    \includegraphics[width=1\textwidth]{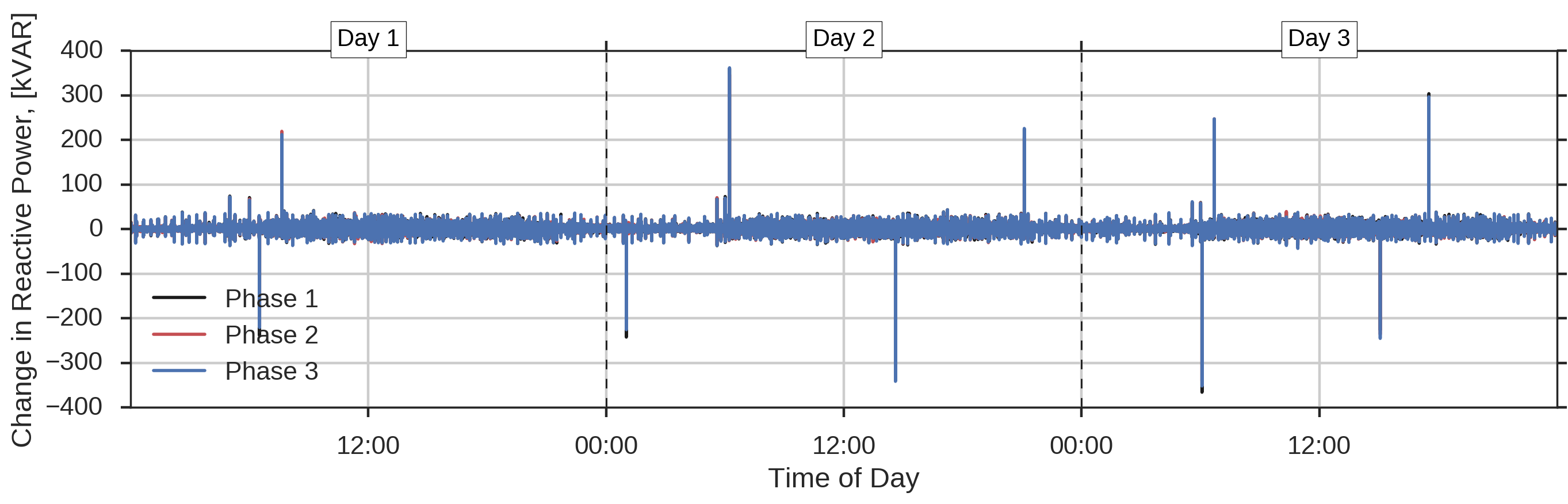}
    \caption{One-second change in the observed reactive power $Q_{PMU_1,t}$}
    \label{fig:deltaReactive}
\end{figure}

We identify instantaneous changes in the reactive power demand to detect the capacitor bank switching events.
For such events, we use the raw $\mu$PMU$_1$ measurements obtained at 120 Hz. 
We calculate differences in measured reactive power once every second using interval averages.   
In Figure~\ref{fig:deltaReactive}, we plot the difference of measured reactive power. These capacitor bank changes can be detected by visual inspection.  

We use a threshold-based detection algorithm to automatically detect capacitor bank switching. Similar algorithms have been used in non-intrusive load monitoring studies~\cite{berges2011user} to identify state changes in devices. In this paper, we propose the following algorithm to detect and account for capacitor bank switching events:

\begin{algorithm}[H]
  compensation=0 \;
  \While{True}{
  $\Delta Q_t=Q_{PMU_1,t-1}-Q_{PMU_1,t}$\;
  \If{$|\Delta Q_t| \geq 90$ kVAR per phase}{
   compensation=$\Delta Q_t+compensation$\;
   }
  \If{$|compensation| < 90$ kVAR per phase}{
  compensation=0\;
  }
  $Q^{filtered}_{PMU_1,t}=Q_{PMU_1,t}+compensation$\;
 }
 \caption{Capacitor Bank Switching Detection and Compensation Algorithm}
 \label{alg:CapBankDetection}
\end{algorithm}
In Algorithm~\ref{alg:CapBankDetection}, the $compensation$ term stores the amount of compensation to reverse the impact of the capacitor bank on the reactive power. We have used a threshold of 90 kilovolt-amperes reactive (kVAR) per phase to detect and label a change in reactive power as capacitor bank reactive power injection. There are various ways to choose an appropriate threshold: using knowledge on the capacitor bank characteristics of the particular circuit, or examination of change of reactive power values using historical measurements---similar to Figure~\ref{fig:deltaReactive}. This detection algorithm is well suited to the discrete switching associated with fixed-sized capacitor banks but would fail to account for more variable reactive power support from static var compensators or similar devices. These devices would offset the measured reactive power demand and affect the performance of the proposed model.

\begin{figure}
    \centering
    \includegraphics[width=1\textwidth]{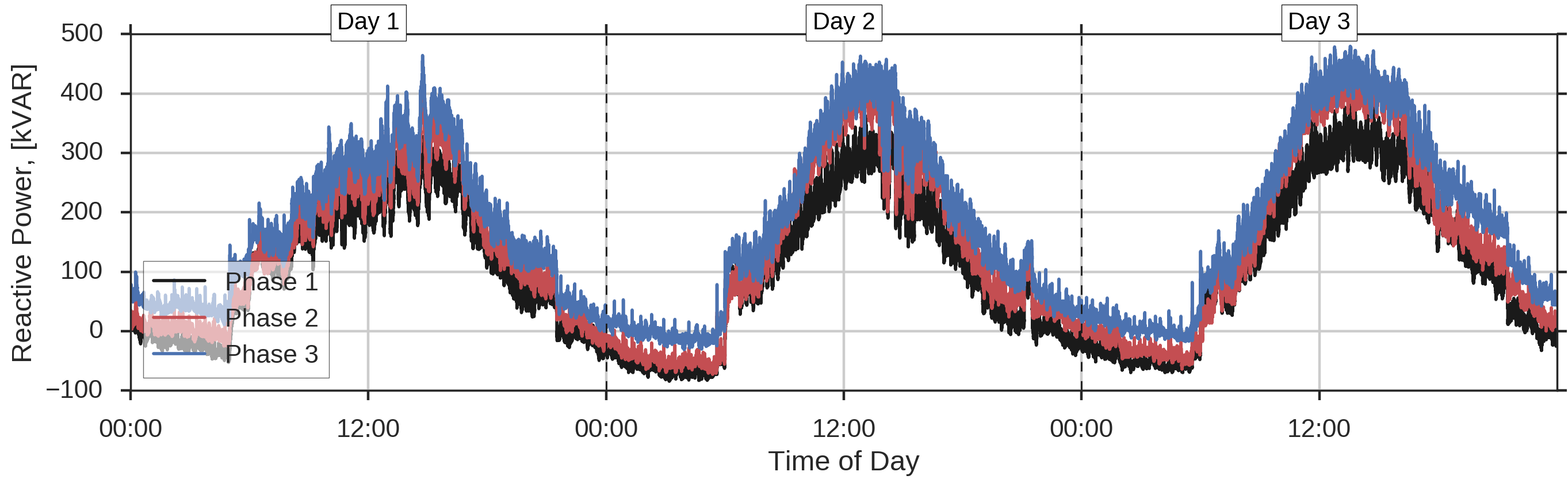}
    \caption{$\mu$PMU$_1$ reactive power corrected for capacitor bank action}
    \label{fig:capbankAfter}
\end{figure}

\begin{figure}
    \centering
    \includegraphics[width=1\textwidth]{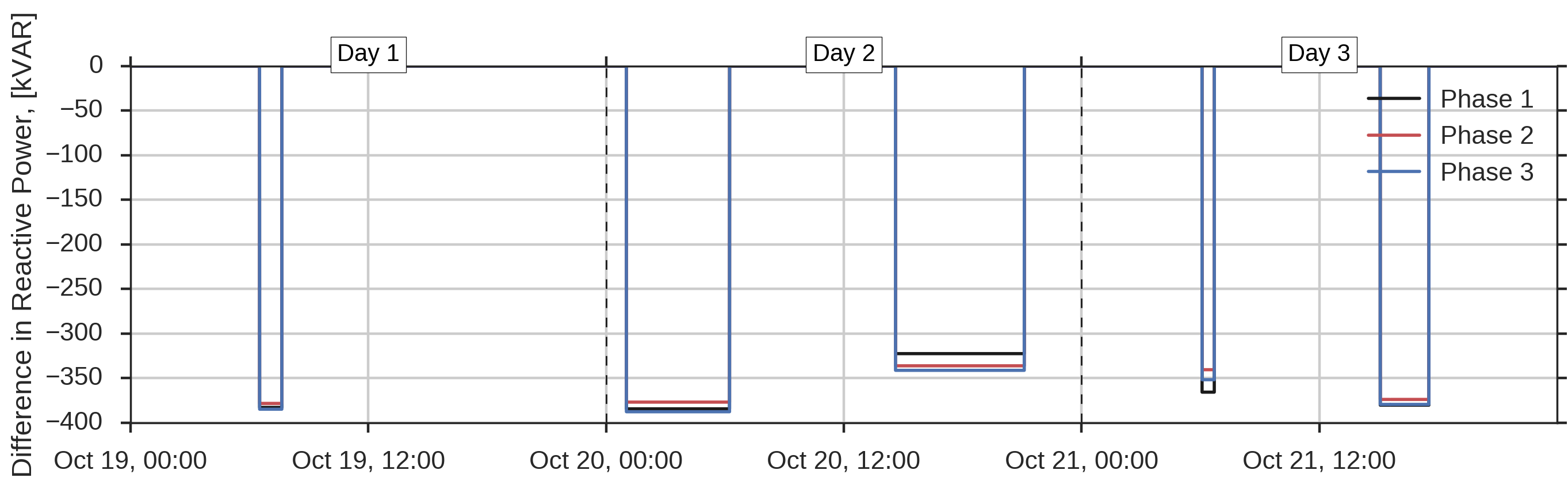}
    \caption{$\mu$PMU$_1$ reactive power difference between the corrected and the original measurements}
    \label{fig:ReactivePowerDifferences}
\end{figure}

Figure~\ref{fig:capbankAfter} shows the reactive power values after the compensation for capacitor bank actions. Figure~\ref{fig:ReactivePowerDifferences} shows the difference between the reactive power values before and after compensation. Specifically, it shows the compensation term defined in Algorithm~\ref{alg:CapBankDetection} in time.

\subsection{Solar Disaggregation Using Load Power Factor}
\begin{figure}[h]
    \centering
    \includegraphics[width=0.4\textwidth]{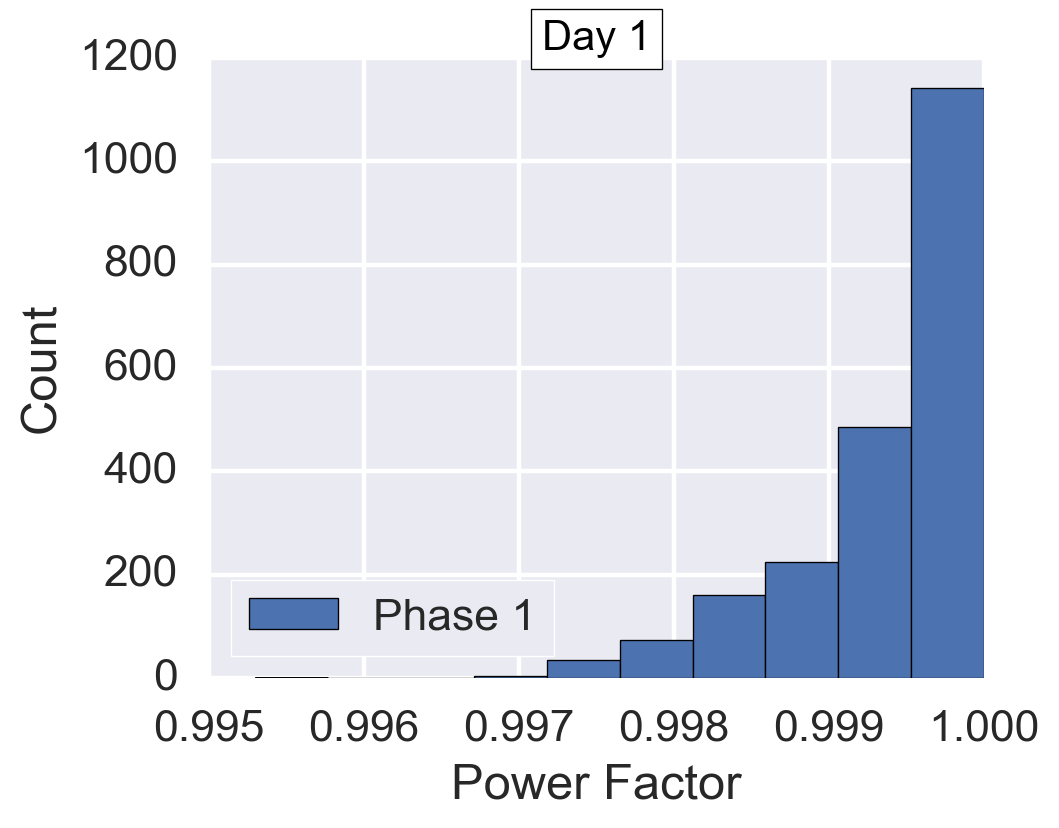}
    \caption{Sample distribution of night power factor}
    \label{fig:LoadPowerFactor}
\end{figure}

After removing the effects of the capacitor bank switching, we propose to use measurements of load power factor obtained overnight to represent the load power factor---neglecting system losses. Figure~\ref{fig:LoadPowerFactor} shows the distribution of load power factor values estimated for Phase 1 using measurements from Day 1 between 12 AM and 5 AM. The median $cos{\Phi^{i}_{PMU,t}}$ where $t \in [12AM,5AM]$ value will be used as a proxy load power factor for disaggregation purposes. We refer to the time invariant estimated power factor of load as $cos{\hat{\Phi}^{i}_{Load}}$. Hence, the $A$ matrix can be rewritten as:
\begin{equation}
\mathbf{A}=
    \begin{bmatrix}
    cos{\hat{\Phi}^{i}_{Load}} & -1\\
    sin{\hat{\Phi}^{i}_{Load}} & 0
    \end{bmatrix}
\end{equation}
For each phase using estimated power factor values and observed measurements from the $\mu$PMU$_1$, one can estimate $S^{i}_{PV,t}$ and $S^{i}_{Load,t}$ using Equation~\eqref{eq:5}:
\begin{equation}
   \mathbf{A}
    \times 
    \begin{bmatrix}
    |S^{i}_{PMU,t}| cos{\Phi^{i}_{PMU,t}}\\ 
    |S^{i}_{PMU,t}| sin{\Phi^{i}_{PMU,t}} 
    \end{bmatrix}
    =
    \begin{bmatrix}
    P^{i}_{PMU,t}\\ 
    Q^{i}_{PMU,t} 
    \end{bmatrix}
    \label{eq:5}
\end{equation}

\begin{table}[]
    \centering
    \begin{tabular}{c|c|c c c|c c c|c c c}
    Case && \multicolumn{3}{c|}{Day 1} & \multicolumn{3}{c|}{Day 2}  & \multicolumn{3}{c}{Day 3} \\
    \hline
     &Phase, $p$&1&2&3&1&2&3&1&2&3\\
    \cline{2-11}
    \multirow{2}{*}{I}&$cos{\hat{\Phi}^{PF}_{Load}}$ & .999 & .995 & .996 & .999 & .997 & .998 & .998 & .999 & .999  \\
    & $cos{\hat{\Phi}^{i}_{PV}}$  & \multicolumn{3}{c|}{-1} &\multicolumn{3}{c|}{-1} &\multicolumn{3}{c}{-1} \\
    \cline{2-11}
    \multirow{2}{*}{II}&$cos{\hat{\Phi}^{PF}_{Load}}$ &\multicolumn{3}{c|}{.97} &\multicolumn{3}{c|}{.97} &\multicolumn{3}{c}{.97} \\
    & $cos{\hat{\Phi}^{i}_{PV}}$  & \multicolumn{3}{c|}{-1} &\multicolumn{3}{c|}{-1} &\multicolumn{3}{c}{-1} \\
    \end{tabular}
    \caption{Estimated and assumed power factor values used in this study.}
    \label{tab:MedianPFLoad}
\end{table}

The estimated median load power factors and the assumed $PV$ power factors are given in Table~\ref{tab:MedianPFLoad} for each day and phase. Case I refers to the estimated power factor via measurements obtained during nighttime. In Case II, we reverse-calculate a power factor based on the measured PV generation value at 2 PM, Day 1, and apply it to all three days. This is to further investigate the time-invariant power factor assumption, and the impact of its value on the estimation performance. 

\begin{figure}[h]
    \centering
    \includegraphics[width=1\textwidth]{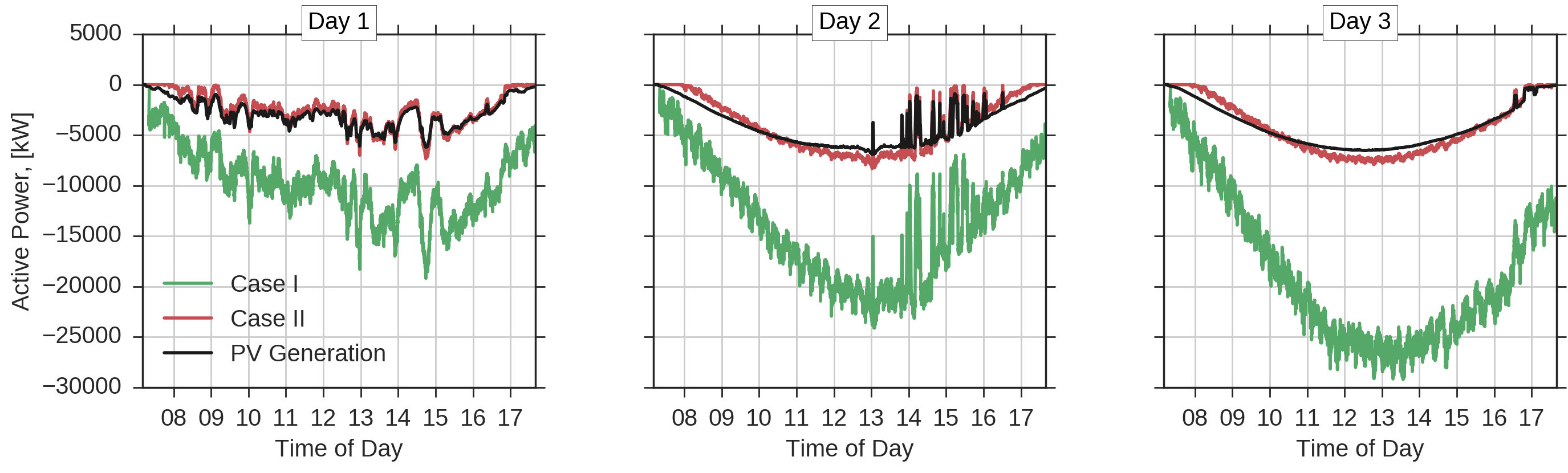}
    \caption{Estimated generation using learned and sample PF}
    \label{fig:pfbe_results}
\end{figure}

Figure~\ref{fig:pfbe_results} shows the estimated generation corresponding to each of the cases outlined in Table~\ref{tab:MedianPFLoad}. Case I shows a poor performance due to various assumptions on power factors of load and PV generation. Case II results suggest that there exists a time-invariant load power factor that can significantly improve the estimation performance over Case I. 

\begin{figure}
    \centering
    \includegraphics[width=1\textwidth]{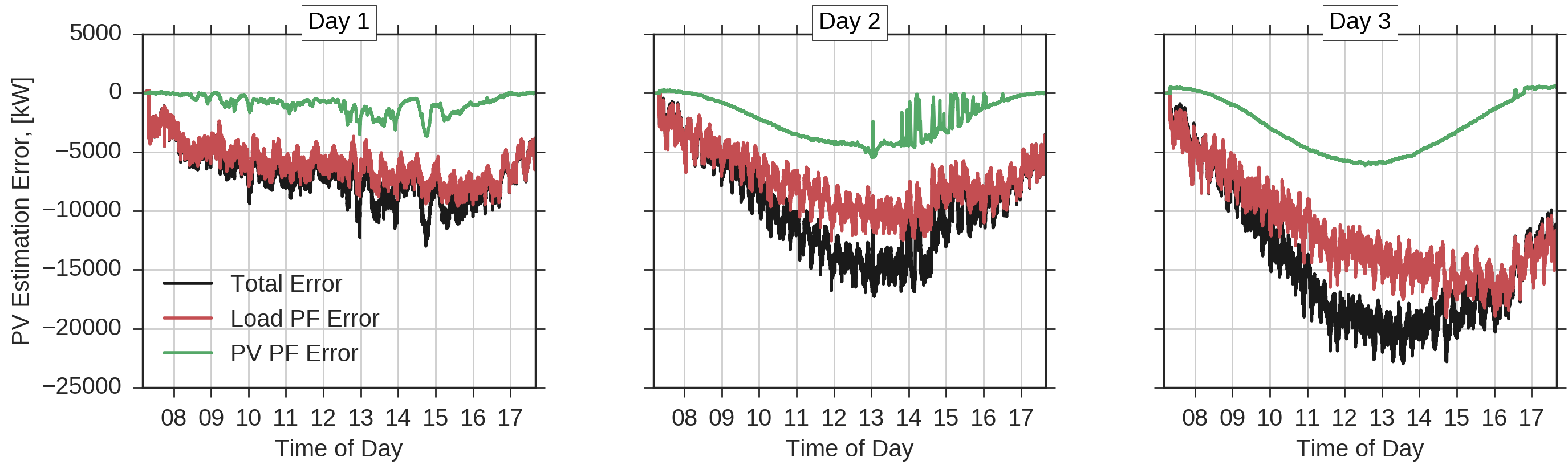}
    \caption{Contribution of each assumption made in PFBE to Case I errors.}
    \label{fig:PFBEErrors}
\end{figure}

To understand the contribution of each assumption made in Case I to the overall estimation error, we break down the contribution of each assumption to the estimation performance. The availability of $\mu$PMU$_2$ measurements makes it possible to compensate individually for each assumption, and reestimate the PV generation, making a single assumption at a time to understand its contribution to the total error. 
Specifically, in Figure~\ref{fig:PFBEErrors}, we show (i) total error: the difference between the real and estimated Case I active generation results, (ii) PV PF error: obtained by using the real load PF values during the estimation period and only making the constant PV power factor assumption (i.e., only active power generation), and (iii) load PV PF error: obtained by using the real PV PF values during the estimation period and only making the constant load power factor assumption (i.e., load power factor estimated over night). It is possible to see that the the constant load PF estimated using measurements overnight is the main reason behind the poor performance of Case I. Assuming a PV power factor of $-1$ also contributes to the overestimation, albeit to a lesser extent. In fact, Figure~\ref{upmu2reacpower} already shows that the PV site consumes reactive power, with a peak consumption $\approx$ 100 kVAR/phase. This reactive power consumption was attributed to the load, and it was responsible for $\approx$ 25\% of the overestimation at its peak contribution.  

In the next sections, building on the results of the PFBE, we propose two methodologies in which we estimate PV and load power factors using daytime measurements. In Section~\ref{sec:LE}, we propose a linear estimator for load and PV that relaxes the assumption that PV systems have a unity power factor, and that a nighttime load power factor is representative of a daytime load power factor. In Section~\ref{sec:CSSE}, we employ contextually supervised source separation, which also allows for variations in load and solar power factor throughout the day. These techniques seek to estimate a power factor closer to that of Case II in Figure~\ref{fig:pfbe_results}. Expanding the PFBE to remove the constant PV power factor is particularly beneficial, as distributed generation is expected to participate in reactive power control in the future \cite{cagnano2011online,arnold2015extremum}. 


\section{Linear Estimator}
\label{sec:LE}
Although the proposed PFBE is computationally inexpensive, it transfers all the variability in reactive power to the load. It also does not include any information on the solar generation. Such information is generally easy to obtain using, for example, a clear sky radiance model, a nearby weather station, or a nearby PV system that monitors and readily reports power generation.
As an initial step towards obtaining a more representative load behavior, we propose a linear model to estimate PV generation $P^{t}_{PV}$ and the load $P^{t}_{Load}$ in the system using the aggregate measurement $P^{t}_{PMU_1}$. Similar to Equation~\eqref{eq:ComplexPower}, we express $P^{t}_{PMU_1}$ as follows:
\begin{equation}
    P^{t}_{PMU_1}=P^{t}_{Load}+P^{t}_{PV}
    \label{eq:ActivePower}
\end{equation}

\noindent Then, we propose to model $P^{t}_{PV}$ using measured generation from a nearby PV system (i.e., the irradiance proxy obtained from the CE-CERT PV system), $\phi^{t}$, where $C_{eff}$ is the coefficient on the generation from the nearby system (i.e., an effective $PV$ capacity) and $\epsilon_{PV}$ is the error term as follows:

\begin{equation}
\label{e:PVLM}
    P^{t}_{PV}= C_{eff}\phi^{t} + \epsilon_{PV}, \hskip14pt \forall t\text{ where }\phi^t > 0
\end{equation}

\noindent For load, $P^{t}_{Load}$, we obtain a similar model. Specifically, we linearly model load as a function of reactive power measured at $\mu$PMU$_1$, $Q^{t}_{PMU_1}$, and an intercept term $R$. The coefficient $k_{eff}$ of $Q^{t}_{PMU_1}$ serves as an average term that captures the effective relationship between the active and reactive power.\footnote{Here, we acknowledge that $cotan(\Phi)=k_{eff}$, where $\Phi$ is the angle between voltage and current phasors.} The intercept, $R$, represents an average constant load in the system with resistive character (i.e., independent of $Q^{t}_{PMU_1}$) or loads who present themselves as so, due to their reactive power component being masked by fixed capacitor banks, and $\epsilon_{Load}$ is the error in load predictions.\\
\begin{equation}
\label{e:LoadLM}
    P^{t}_{Load}=k_{eff}Q^{t}_{PMU_1} +  R + \epsilon_{Load}, \hskip14pt \forall t\text{ where }\phi^t > 0  
\end{equation}

In our current estimation setting, we do not have access to $P^{t}_{Load}$ or $P^{t}_{PV}$. However, we observe the aggregate measurements $P^{t}_{PMU_1}$. Hence, using Equation~\eqref{eq:ActivePower} we combine the models presented above as follows:
\begin{equation}
    P^{t}_{PMU_1}=k_{eff}Q^{t}_{PMU_1}+C_{eff}\phi^{t}+R+\overbrace{\epsilon_{PV}+\epsilon_{Load}}^{\epsilon_{Total}}, \hskip14pt \forall t\text{ where }\phi^t > 0 
    \label{eq:MultiRegression}
\end{equation}
Since both ${P}^t_{Load}$ and ${P}^t_{PV}$ have their own noise terms, we cannot recover these signals directly using OLS.
However, we can recover the coefficients $C_{eff}$ and ${k}_{eff}$ through OLS. 
\begin{figure}
\centering
\includegraphics[width=0.5\textwidth]{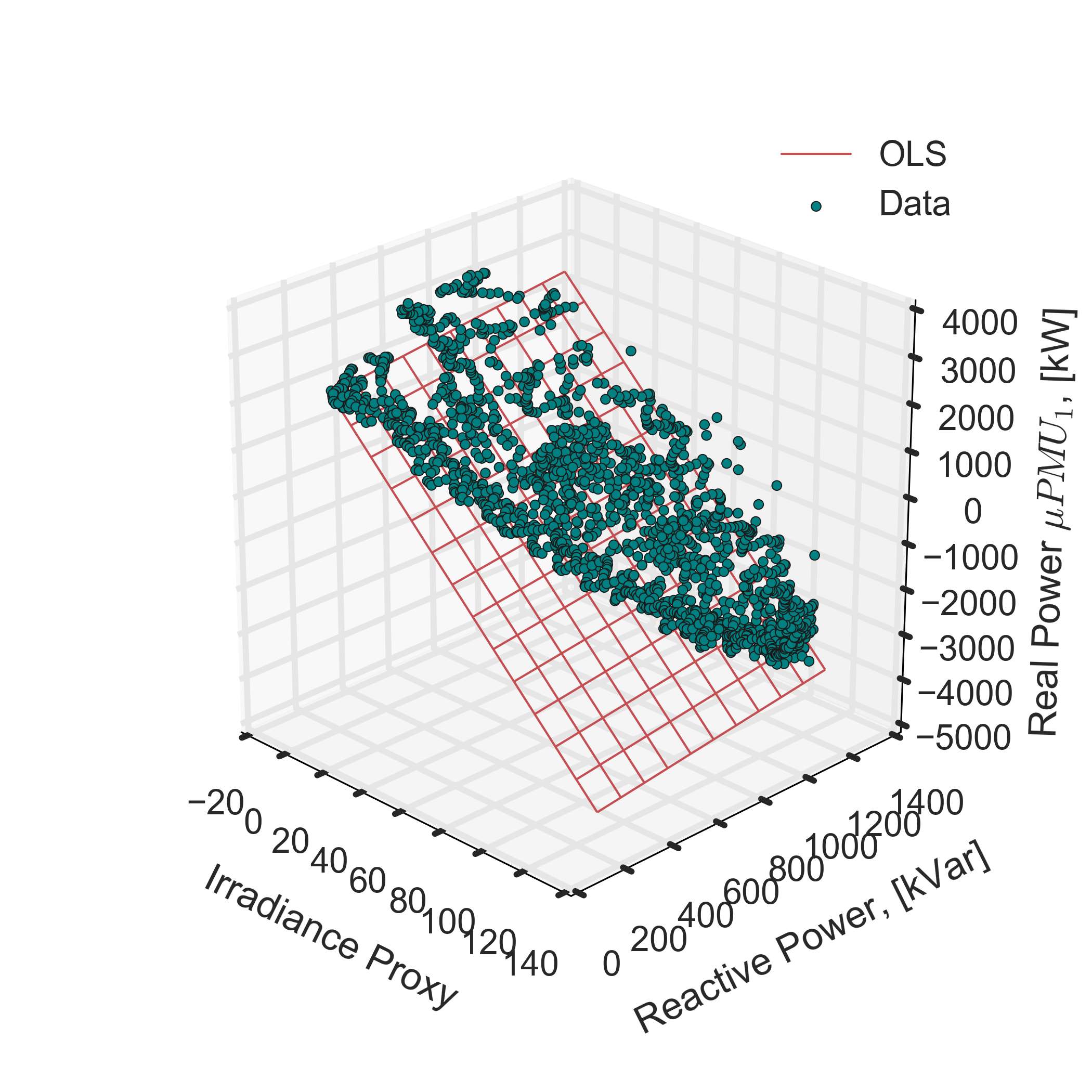}
\caption{The data points and corresponding OLS results for the model in Equation~\eqref{eq:MultiRegression}}
\label{fig:multi_regression_ols}
\end{figure}

In Figure~\ref{fig:multi_regression_ols}, we show the resulting plane represented by the multiple linear regression model in Equation~\eqref{eq:MultiRegression}. Table~\ref{tab:MultiRegressionCoeff} includes the identified coefficients as well as the $R^2$ metric for the proposed model in Equation~\eqref{eq:MultiRegression}. It is important to note that the negative $C_{eff}$ value is due to the sign convention used in this paper (i.e., positive for consumption behind the substation).


\begin{table}[h]
    \centering
    \begin{tabular}{c|c}
    Coefficients & LE in~\eqref{eq:MultiRegression}\\
    \hline 
         \multirow{2}{*}{$R$ (intercept)} & 2193.556$^{***}$\\
         &(81.402)\\
         \multirow{2}{*}{$k_{eff}$} & 1.05$^{***}$\\
         & (0.161) \\
          \multirow{2}{*}{$C_{eff}$} & -47.454$^{***}$\\ 
          &(1.015)\\
         \hline
         Adjusted $R^2$ & 0.749\\
         Number of observations & 1896\\
    \end{tabular}
    \caption{Regression coefficients for the model in Equation~\eqref{eq:MultiRegression}. Standard errors are reported in parentheses. The asterisks (*, **, ***) indicate significance at the 90\%, 95\%, and 99\% level, respectively.}
    \label{tab:MultiRegressionCoeff}
\end{table}
To reconstruct the PV generation and the load using the coefficients given in Table~\ref{tab:MultiRegressionCoeff}, we need to make assumptions on how to distribute $\epsilon_{Total}$ to $\epsilon_{Load}$ and $\epsilon_{PV}$.
Given the size of the PV system studied here and the generally high variability of PV generation compared to load, we expect that $\epsilon_{Load}\ll\epsilon_{PV}$; thus we assume $\epsilon_{PV} \approx \epsilon_{Total}$. Equation~\eqref{eq:MLR_estimation} applies this assumption to obtain an estimate of $P^{t}_{Load}$, which is denoted as $\hat{P}^t_{Load}$, and similarly, an estimate of ${P}^t_{PV}$, $\hat{P}^t_{PV}$, by assigning all residuals from the regression to PV generation.

\begin{equation}
\label{eq:MLR_estimation}
\begin{aligned}
    &\hat{P}^t_{Load}= {k}_{eff} Q^{t}_{PMU_1} +  R\\
    &\hat{P}^t_{PV}=P^{t}_{PMU_1}-\hat{P}^t_{Load}\\
\end{aligned}
\end{equation}

\begin{figure}
\subfloat[$\hat{P}^t_{Load}$ estimated using the LE (i.e. Equation~\eqref{eq:MLR_estimation}) \label{fig:reg_estimated_load}]{    \includegraphics[width=1\textwidth]{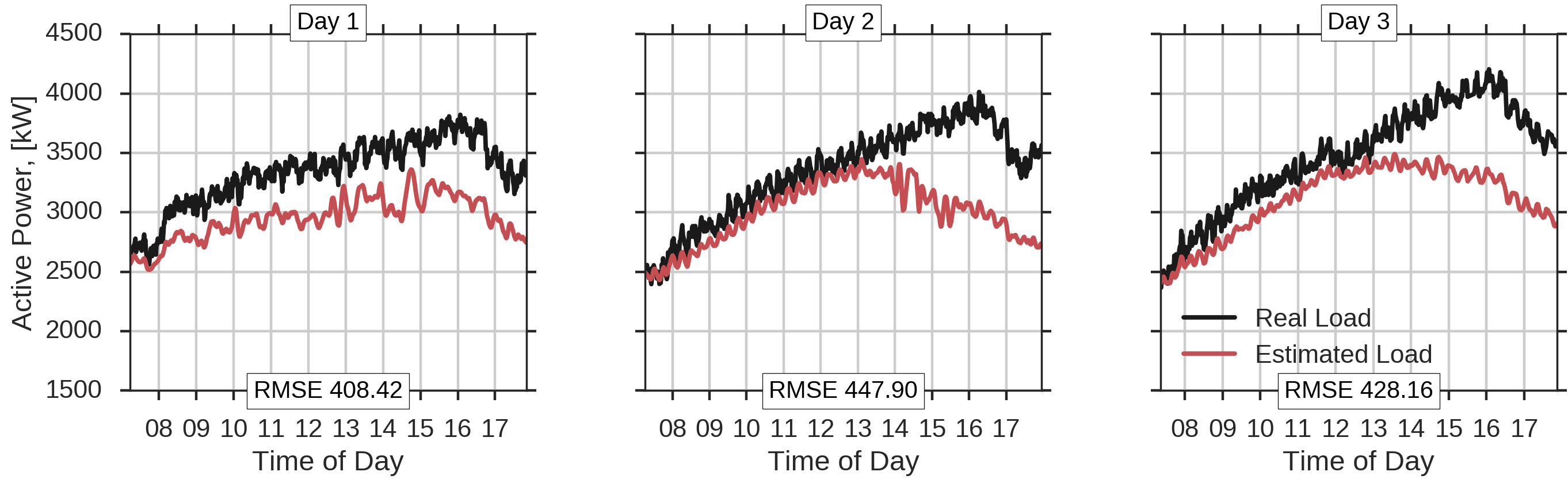}}
\hfill
\subfloat[$\hat{P}^t_{PV}$ estimated using the LE (i.e. Equation~\eqref{eq:MLR_estimation}) \label{fig:reg_estimated_solar}]{  \includegraphics[width=1\textwidth]{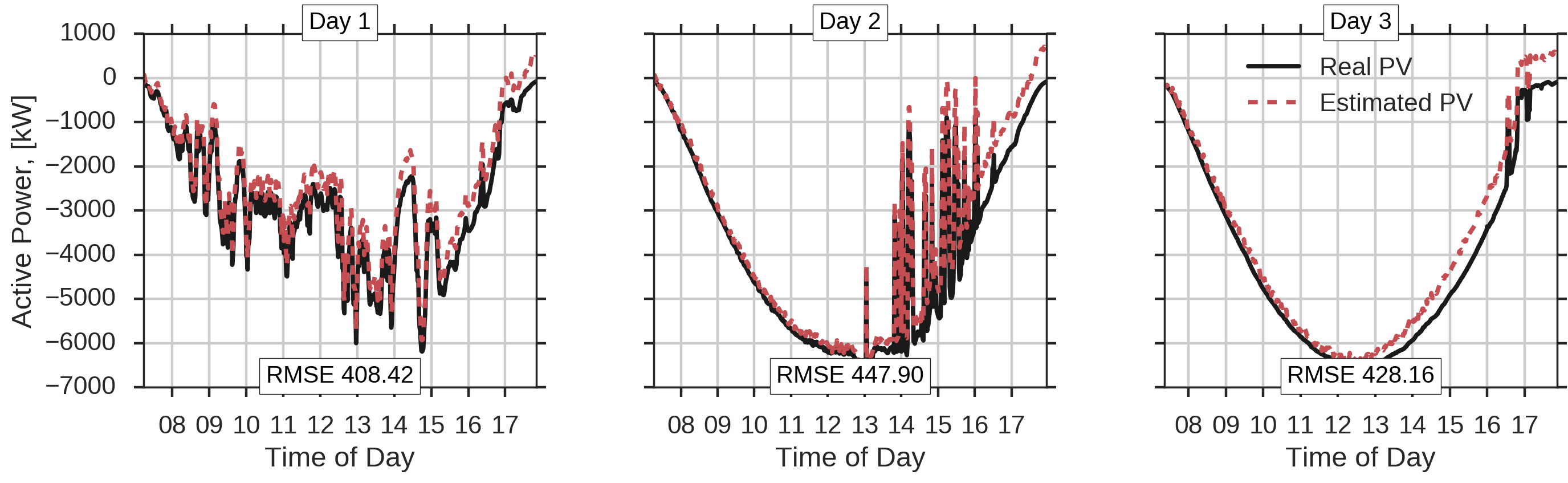}}
\label{fig:my_label}
\caption{Disaggregation results for the linear estimator}
\end{figure}

Using the proposed set of equations in~\eqref{eq:MLR_estimation}, one can obtain $\hat{P}^t_{Load}$ and $\hat{P}^t_{PV}$. The resulting $\hat{P}^t_{Load}$ and $\hat{P}^t_{PV}$ as well as corresponding RMSE values between the estimated and measured load and generation are presented in Figures~\ref{fig:reg_estimated_load} and~\ref{fig:reg_estimated_solar}, respectively. Since we estimate the solar generation based on predictions of load in Equation~\eqref{eq:MLR_estimation}, the RMSE values for both load and solar estimations are identical for all days. In Figure~\ref{fig:reg_estimated_solar}, it is possible to see that the linear estimator outperforms the PFBE (Case I) in estimating the solar generation at the feeder. However, the assumption on the errors does not take the true performance of individual load and PV models into consideration; rather, it asserts 100\% confidence on the load model. This limits the applicability of LE to other systems where the assumption on the errors might not hold true.

In the following section, we employ contextually supervised source separation to distribute the overall error term, $\epsilon_{Total}$, into both the load and PV error terms, rather than assigning it entirely to PV.   

\section{Contextually Supervised Generation Estimation}
\label{sec:CSSE}
To expand the proposed linear regression setting, we explicitly represent reconstructed signals and capture the errors systematically using contextually supervised source separation.
In its general form, contextually supervised source separation is introduced in~\cite{wytock2013contextually} as a single-channel source separation methodology. As opposed to supervised settings that require true signal separations~\cite{kolter2010energy}, and unsupervised settings~\cite{kim2011unsupervised}, which result in arbitrarily many solutions, it leverages contextual supervision to disaggregate the source signal into correlated component signals. In~\cite{wytock2013contextually}, the authors apply this methodology to the energy disaggregation problem. 

For $L$ many unknown signals that are of interest, we observe the aggregate signal, $Y_{agg}$ such that
\begin{equation}
Y_{agg}=\sum_{i=1}^{L} Y_{i}
\end{equation}
where $Y_{i}$ represents an individual unknown signal.
Assuming we can represent the individual load with a linear model (i.e., $Y_i ~X_i\Theta_i$), we introduce the general contextually supervised learning problem as follows:
\begin{equation}
\label{eq:se:CSSE}
\begin{aligned}
 &\underset{Y_{i},\Theta_{i}}{\text{minimize}}
&&\{\mathbf{\alpha_i}\ell_i((Y_{i}-X_{i}\Theta_{i})+\eta_i g_i(Y_{i})+ \gamma_i h_i(\Theta_i) \} \\
& \text{subject to}
&& Y_{agg}=\sum_{i=0}^L Y_{i}\\
\end{aligned}
\end{equation}
\noindent where $\ell$ represents a function that penalizes the difference between the reconstructed signal and the corresponding linear model. The function $g$ represents a penalty function that captures additional contextual information on individual signals, such as smoothness. Finally, the function $h$ represents a regularization term on the model parameters to prevent overfitting. In~\cite{wytock2013contextually}, the authors discuss the differences between using various combinations of $\ell$, $g$, and $h$ in the context of energy disaggregation. The authors also discuss the special case in which only $\ell_i$ function is used in the objective function as an $\ell-2$ norm. This boils the proposed methodology down to an ordinary linear regression in which the $Y_{agg}$ term is regressed by the $\Theta_i$ parameters.

In the next section, we first formulate the solar disaggregation problem at hand as a generic contextually supervised source separation problem. We then boil it down to the special case proposed by~\cite{wytock2013contextually} due its similarities with LE. 
\subsection{Problem Formulation} 
\label{sec:formulation}
Assuming that proxy irradiance, $\phi^t$, and $\mu$PMU$_1$ measurements are available to the estimator, we cast the case-specific disaggregation problem as the following optimization for all t where $\phi^t > 0$:
\begin{equation}
\begin{aligned}
 &\underset{P_{Load},P_{PV},k_{eff},C_{eff} }{\text{minimize}}
&&\{\mathbf{\alpha}\ell_1(P^{t}_{Load}-(k_{eff} Q^{t}_{PMU_1}+R))+\mathbf{\beta}\ell_2(P^{t}_{PV}-C_{eff} \phi^{t})\\
&&&+g_1(P^t)+g_2(P^t_{Load})\\
&&&+h_1(C_{eff})+h_2(k_{eff})\} \\
& \text{\:\:\:\:\:\:\:\:\: subject to}
&& P^{t}_{PMU_1}=P^{t}_{Load}+P^{t}_{PV} \\
\end{aligned}
\label{eq:main_problem}
\end{equation}
\noindent In Equation~\eqref{eq:main_problem}, we represent the load $P^{t}_{Load}$ as a linear function of reactive power measured at PMU$_1$ $Q^{t}_{PMU_1}$, and the PV generation $P^{t}_{PV}$ as a linear function of a solar irradiance proxy measurement $\phi^{t}$. Similar to the LE, we refer to the reactive power coefficient as the \emph{effective power factor} $k_{eff}$, and the solar irradiance proxy measurement coefficient as the effective solar capacity $C_{eff}$. The $\alpha$ and $\beta$ parameters are weighting factors. The loss functions $\ell_1$ and $\ell_2$ are penalizing the differences between the reconstructed solar generation and load, and their linear representation is modeled using $\phi^{t}$ and $Q^{t}_{PMU_1}$, respectively.

The problem presented in Equation~\eqref{eq:main_problem} is the main disaggregation problem addressed in its most generic form. In this paper, we are interested in the specific case where we only use second norms for the $\ell_i$ terms only. Hence, we are interested in the following problem:
\begin{equation}
\begin{aligned}
 &\underset{P_{Load},P_{PV},k_{eff},C_{eff} }{\text{minimize}}
&&\{\mathbf{\alpha}||P^{t}_{Load}-(k_{eff} Q^{t}_{PMU_1}+R)||^2+\mathbf{\beta}||P^{t}_{PV}-C_{eff} \phi^{t}||^2\}\\
& \text{\:\:\:\:\:\:\:\:\: subject to}
&& P^{t}_{PMU_1}=P^{t}_{Load}+P^{t}_{PV} \\
\end{aligned}
\label{eq:reduced_problem}
\end{equation}

In~\cite{wytock2013contextually}, the authors show that the specific case presented in Equation~\eqref{eq:reduced_problem} and the multiple linear regression strategy presented in Equation~\eqref{eq:MultiRegression} results in the identical model parameters $C_{eff}$, $k_{eff}$ and $R$ for all cases where $\alpha=\beta$. In~\ref{sec:alpha_beta}, we show that this is true for any non-zero $\alpha$ and $\beta$ values.

As discussed in Section~\ref{sec:LE}, each component of the proposed linear models has its own error term, and we cannot expect to recover the true $P_{Load}$ and $P_{PV}$. We can only obtain the truthful values for $C_{eff}$ and $k_{eff}$ under an ordinary least squares (OLS) setting. However, our main focus is to obtain accurate reconstructed signals $P_{Load}$ and $P_{PV}$. 

Unlike the LE, instead of making a simple assumption on the magnitude of the errors for each model, we propose to estimate weights multiplying these error values---$\alpha$ and $\beta$---exogenously from day and night observations. Specifically, we only assume that the individual errors are Gaussian, and independent and identically distributed (i.i.d.). We refer to the error term of the linear model of load as $\epsilon_{Load}$ and of PV as $\epsilon_{PV}$. The following holds for the aggregate error, given that $\epsilon_{Load}$ and $\epsilon_{PV}$ are i.i.d:

\begin{equation}
Var(\epsilon_{Total}) \approx Var(\epsilon_{PV})+Var(\epsilon_{Load})
\label{eq:iid}
\end{equation}

To estimate $Var(\epsilon_{Load})$, we leverage nighttime measurements, when solar generation is not available. Specifically, we propose to estimate the variance of the error term $\epsilon_{Load}$ using the following linear model:
\begin{equation}
    P_{PMU_1}=k_{eff} Q^{t}_{PMU_1}+R+\epsilon_{Load}, \hskip14pt \forall t\text{ where }\phi^t \le 0
\label{eq:linear_model_load}
\end{equation}

To obtain the variance in the PV generation, it is possible to use measurements obtained from $\mu$PMU$_2$. However, in reality, we would not have measurements obtained from $\mu$PMU$_2$. Hence, we propose to use the property given in Equation~\eqref{eq:iid} to estimate the variance of PV. Specifically, we first calculate $Var(\epsilon_{PMU_1})$ during daytime periods. Then, assuming the nighttime $Var(\epsilon_{Load})$ is representative of the daily variance in the load, we calculate the variance in PV. 
Formally, to estimate $Var(\epsilon_{PV})$, we first obtain $Var(\epsilon_{Total})$ using the following model used in LE:
\begin{equation}
    P_{PMU_1}=k_{eff} Q^{t}_{PMU_1}+C_{eff} \phi^{t} + R + \epsilon^*_{Total}, \hskip14pt \forall t\text{ where }\phi^t > 0
\label{eq:linear_model_aggregate}
\end{equation}
\noindent We then use the property introduced in Equation~\eqref{eq:iid} to estimate $Var(\epsilon_{PV})$.

To distribute the errors relative to the \emph{goodness} of each model (i.e., to minimize the variance of the weighted average), we multiply the expected error in the objective function by the inverse of its variance. Hence, we define $\alpha^*$ and $\beta^*$ as follows:
\begin{equation}
    \alpha^*=1/Var(\epsilon_{Load}), \hskip14pt \beta^*=1/Var(\epsilon_{PV})
\label{eq:definition_alpha_star}
\end{equation}

\begin{figure}
\centering
\includegraphics[width=0.8\textwidth]{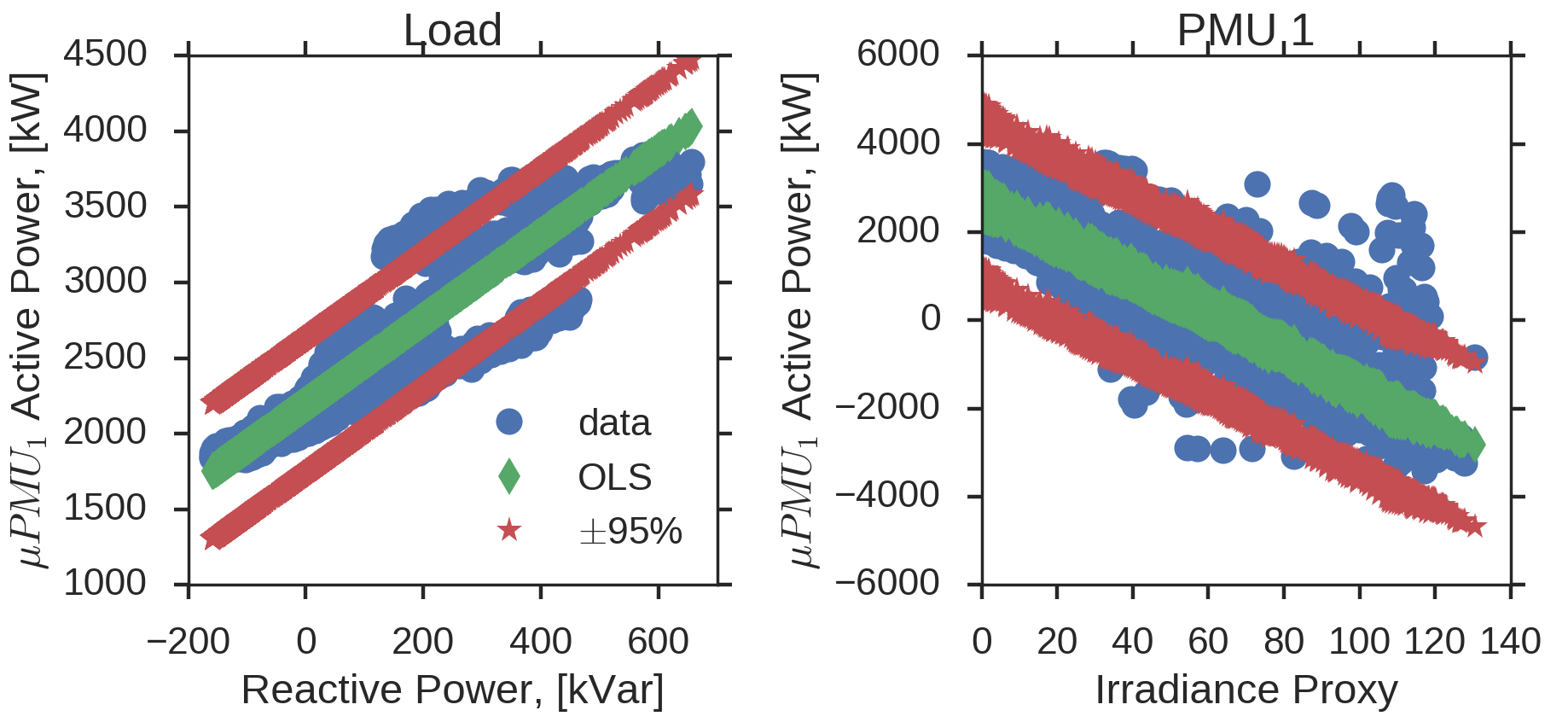}
\caption{Data, the resulting fit, and 95\% confidence intervals for both models introduced in Equations~\eqref{eq:linear_model_load}, and~\eqref{eq:MultiRegression} and~\eqref{eq:linear_model_aggregate}}
\label{fig:variance_ols}
\end{figure}

Figure~\ref{fig:variance_ols} shows the resulting models used to estimate $Var(\epsilon_{Load}$) and $Var(\epsilon_{PV}$). They show the data used in fitting the models, the resulting fit and 95\% confidence intervals for the models introduced in Equations~\eqref{eq:linear_model_load} and~\eqref{eq:linear_model_aggregate}, respectively. The left plot shows the model obtained by regressing the active load in the system using reactive power overnight as given in Equation~\eqref{eq:linear_model_load}. The right plot shows the aggregate model as given in Equation~\eqref{eq:linear_model_aggregate} with respect to solar irradiance. 
The resulting coefficients for both of these models are included in Table~\ref{tab:ResidualRegressionCoeff}. Note that, the results shown in the right column are identical to Table~\ref{tab:MultiRegressionCoeff} and included here for completeness. All the resulting coefficients show significance at 99\%. Given very similar $R$ values for both models, if we assume no network losses, the difference between the estimated $k_{eff}$ values suggest more active power consumption during nighttime given a $Q^{t}_{PMU_1}$ value, suggesting a higher PF during nighttime.  

\begin{table}[h]
    \centering
    \begin{tabular}{c|c|c}
    Coefficients & Model in~\eqref{eq:linear_model_load} & {LE in~\eqref{eq:MultiRegression} and~\eqref{eq:linear_model_aggregate}}\\
    \hline
         \multirow{2}{*}{$R$ (intercept)} & 2194.32$^{***}$ & 2193.556$^{***}$\\
         & (5.448) & (81.402 )\\
         \multirow{2}{*}{$k_{eff}$} & 2.795$^{***}$ & 1.05$^{***}$ \\
         & (0.025) & (0.161) \\
          \multirow{2}{*}{$C_{eff}$} & & -47.454$^{***}$\\ 
          & &(1.015)\\
         \hline
         Adjusted $R^2$ &  0.849 &  0.749\\
         Number of observations & 2295& 1896\\
    \end{tabular}
    \caption{Regression coefficients for the models in Equation~\eqref{eq:linear_model_load}, and Equations~\eqref{eq:MultiRegression} and~\eqref{eq:linear_model_aggregate}. Standard errors are reported in parentheses. The asterisks (*, **, ***) indicate significance at the 90\%, 95\%, and 99\% level, respectively.}
    \label{tab:ResidualRegressionCoeff}
\end{table}



\begin{figure} [h]
\subfloat[Estimated load using CSGE\label{fig:csse_estimated_load}]{  \includegraphics[width=1\textwidth]{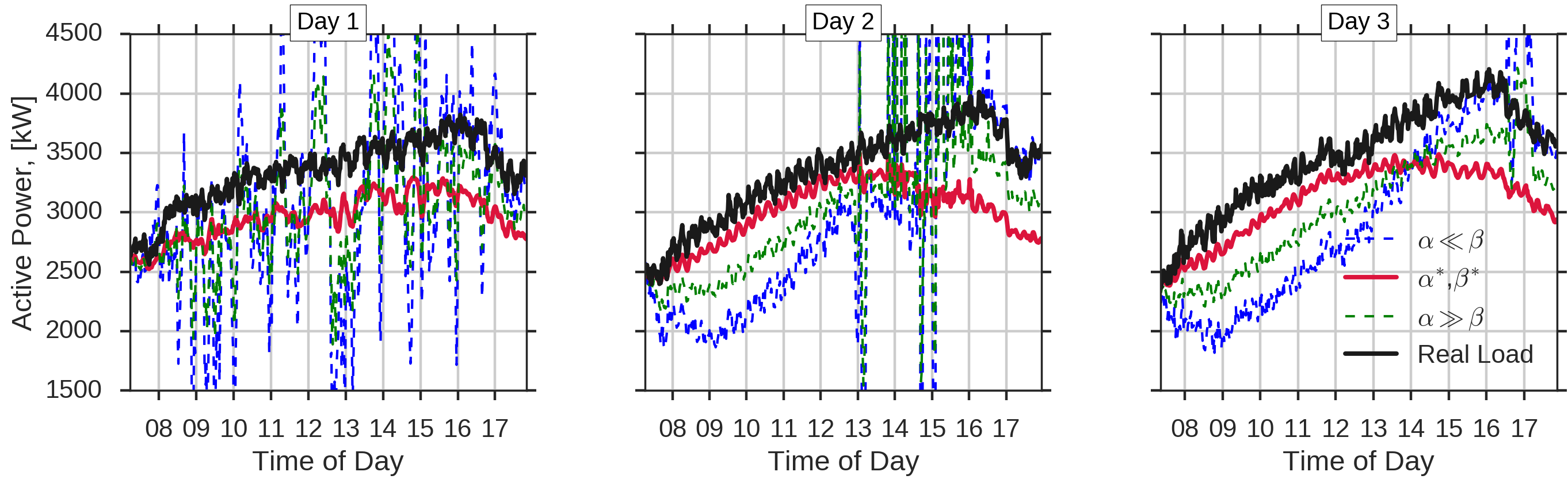}}
\hfill
\subfloat[Estimated solar using CSGE\label{fig:csse_estimated_solar}]{    \includegraphics[width=1\textwidth]{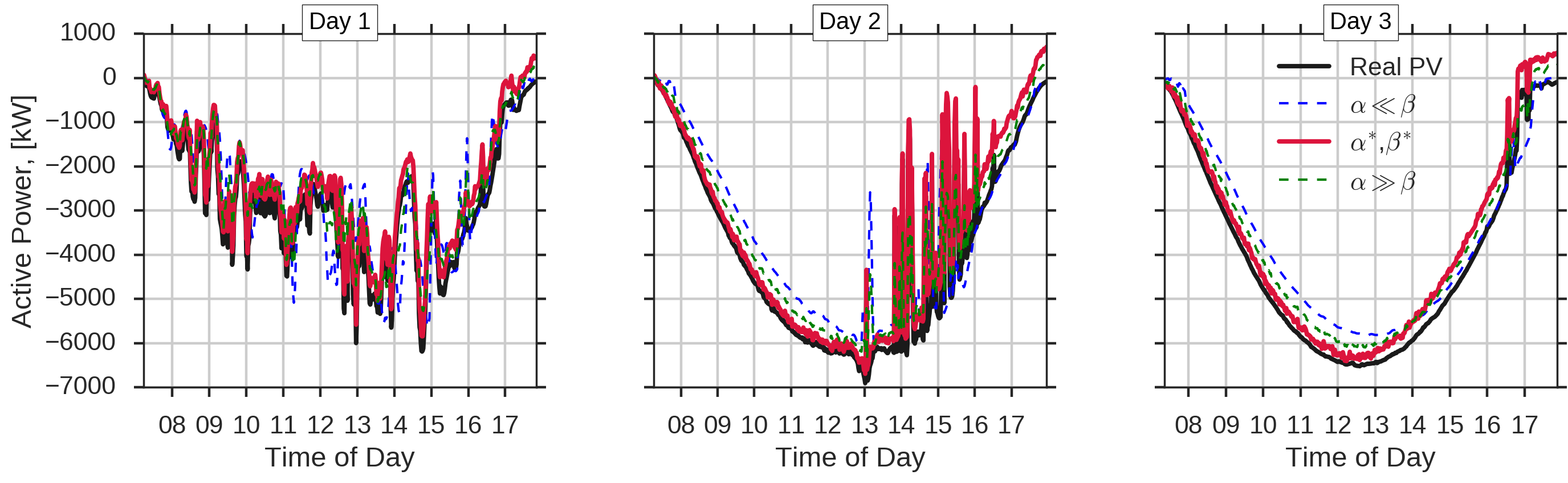}}
\label{fig:CSSE_disaggregation_results}
\caption{Load and solar estimation results using the contextually supervised strategy}
\end{figure}

Figures~\ref{fig:csse_estimated_load} and~\ref{fig:csse_estimated_solar} present the CSGE results for load and PV, respectively. Specifically, we present results for three different cases: $\alpha^*/\beta^*$, $\alpha \ll \beta$, and $\alpha \gg \beta$. For Day 1 and Day 2, it is possible to see that when $\alpha \ll \beta$, the reconstructed load signal shows a very noisy behavior. This is because decreasing $\alpha$ and increasing $\beta$ allows us to enforce the solar representation to be as close as possible to the proxy measurement $\phi^t$. 
In other words, we significantly increase the weighting factor of the deviation of the reconstructed solar generation signal from the linear model that is representing it. This is especially clear for the Hour 13 of the Day 2 results. The opposite is true for $\alpha \gg \beta$. In fact, we should expect the results of $\alpha \gg \beta$ to be very close to the $LE$. This is because in $LE$, we made the assumption that the error in the linear model of the load is negligible in comparison to the error in the PV model. 

\begin{figure}
\includegraphics[width=1\textwidth]{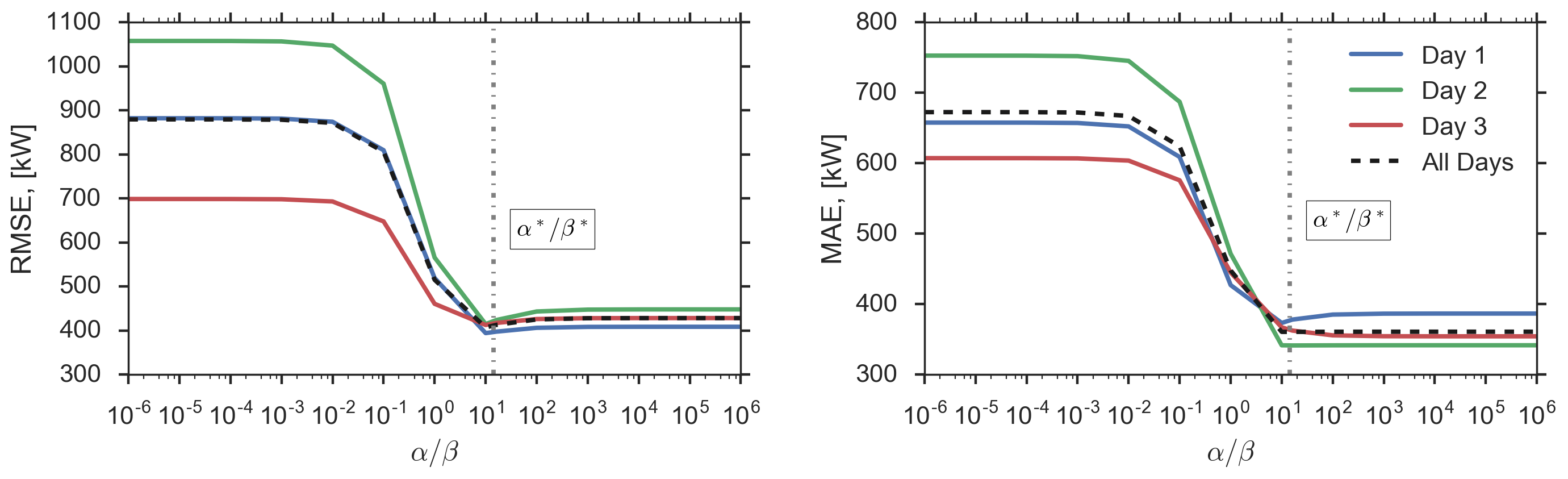}
\caption{The sensitivity of disaggregation accuracy to $\alpha/\beta$.The left plot shows the RMSE between the real and estimated solar generation and load, while the right plot shows the mean absolute error (MAE) between the real and estimated load}
\end{figure}

\begin{figure}
    \centering
    \includegraphics[width=1\textwidth]{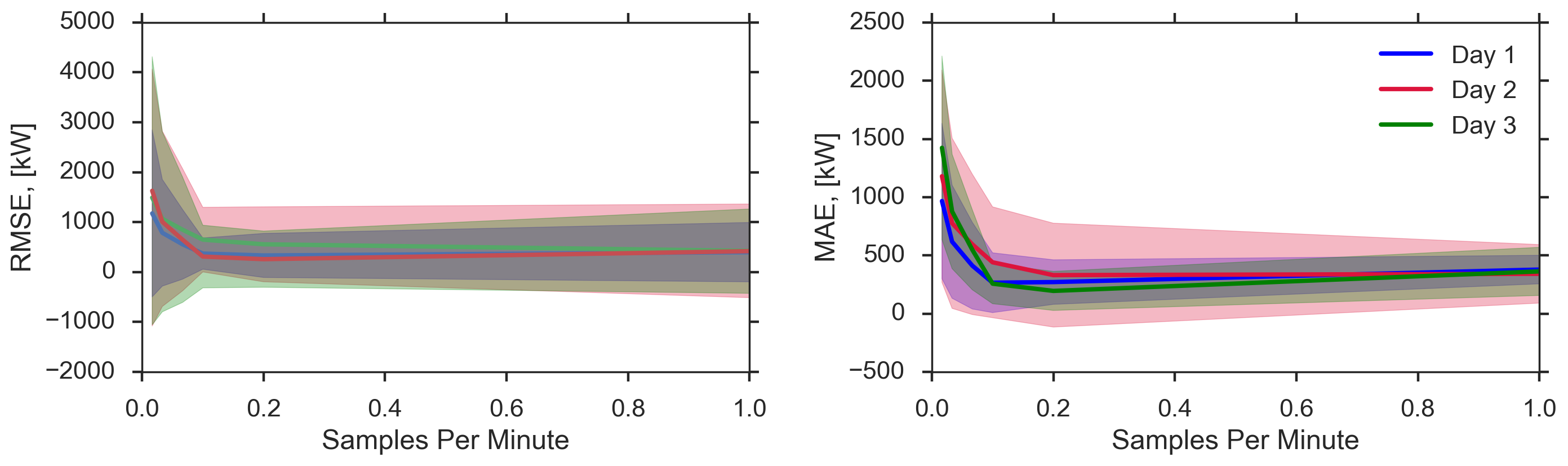}
    \caption{The sensitivity of disaggregation accuracy to sampling rate.The left plot shows the RMSE between the real and estimated solar generation and load, while the right plot shows the MAE between the real and estimated load. The shaded areas correspond to $\pm 2\sigma$ range for the errors.}
    \label{fig:CSSE_MAE_RMSE}
\end{figure}

To further demonstrate the role of $\alpha$ and $\beta$ in disaggregation performance, we evaluate the disaggregation accuracy for $P^{t}_{Load}$ and $P^{t}_{PV}$ for a set of $\alpha/\beta$ parameters, including $\alpha^*/\beta^*$. 
Figure~\ref{fig:CSSE_MAE_RMSE} show the sensitivity of disaggregation accuracy measured by mean absolute error (MAE) and RMSE, respectively. Note that for both MAE and RMSE, for $\alpha/\beta$ values less than the estimated $\alpha^*/\beta^*$ the disaggregation performance improves significantly. For values of $\alpha/\beta$ higher than  $\alpha^*/\beta^*$, the disaggregation performance remains the same. This is consistent with our findings in LE. We conclude that the assumptions made on individual model errors in LE are almost correct. That is, given the size of the PV system studied here, and the generally high variability of PV generation compared to load, our results suggest that $\epsilon_{Load}\ll\epsilon_{PV}$. We also should note that the contextually supervised generation estimation results in a slightly lower reconstructed signal RMSE in comparison to LE, and it also provides us with a strategic way to manage the errors between the linear representations of each signal and their reconstruction.

Having a one-minute sampling rate for both proxy irradiance measurements and the substation measurements makes it possible to study the errors of the proposed methodology with respect to a wide range of measurement sampling rate. Although most smart meters currently only transfer energy measurements, they can also be used to obtain reactive power measurements~\cite{turitsyn2011options}. These measurements can be aggregated at the substation level to be used in generation disaggregation as proposed in this paper. Furthermore, steady-state SCADA units provide reactive power measurements at a 15-minute sampling rate. Hence, we propose to down-sample the inputs to the estimation strategy to evaluate how accurate one-minute PV generation disaggregation results would be using input data with lower sampling rates.

To do that, we create input signals by padding the latest measurement to conduct one-minute disaggregation. We then estimate the $\alpha^*$ and $\beta^*$ values for each sampling rate. We use the contextually supervised generation estimation method to obtain reconstructed signals. We then report average RMSE and MAE values, as well as $\pm 2\sigma$ ranges around these errors. Figure~\ref{fig:CSSE_MAE_RMSE} shows the results. Notice that, for up to 0.2 samples per minute (i.e. a five-minute sampling rate), we observe significant improvement in both average MAE and RMSE and the variance terms. After 0.2 samples per minute, there is no improvement in the RMSE and MAE values. This might be due to the low-pass filter (i.e. five-minute rolling mean) applied to the solar data. The added benefit of one-minute reactive power measurements as opposed to five-minute measurments does not seem to improve the overall performance of the proposed contextually supervised disaggregation methodology when a five-minute rolling mean filter is applied. A detailed analysis of varying input data granularity and filtering is left for future work.   


\section{Conclusions}
In this paper we propose a set of methods to disaggregate PV active power generation from active power consumption measured at a distribution substation or feeder head. We first propose a power factor-based estimator (PFBE), in which we estimate the load's power factor using nighttime measurements, when PV generation is not generating, and we assume that PV's power factor is -1. 
Although the PFBE is computationally inexpensive, its assumptions greatly limit its ability to disaggregate PV. 
Specifically, the assumption of nighttime power factor being representative of the daytime power factor is particularly limiting, and reactive power consumption of the PV plant is not zero. 
In fact, at times, reactive power consumption of the PV plant exceeds 25\% of the load's reactive power consumption.  
These assumptions result in a significant overestimation of active power consumption and, correspondingly, PV generation in the system. 

In order to relax some of the assumptions made in the PFBE, we introduce the linear estimator.
The linear estimator predicts load using reactive power and a constant term, and predicts PV generation using observed generation from a nearby PV system. 
We fit coefficients to each model by adding them together and minimizing the square errors in their prediction of total active power measured at the substation. 
Using the estimated linear models, and an assumption that errors in the load predictions are much less than errors in PV generation predictions, $\epsilon_{Load}\ll\epsilon_{PV}$, we reconstruct the PV generation signal for three days. We observe a significant improvement in performance; however, our assumption on the errors is limiting. In particular, it does not reflect the expected performance of each model. Rather, it assumes an accurate load model and associates all the errors with the PV model. Thus, it limits our ability to tune the disaggregation based on the expected performance of each model.

To better distribute the errors from the aggregate model estimated by LE to the individual models, we propose to use contextually supervised generation separation, CSGE \cite{wytock2013contextually}. 
We extend Wytock and Kolter's original analysis to show that, for the special case $\ell$-2 norm objective functions used here, the optimal solution for model coefficients are the same as those found by the LE, regardless of how each model's prediction errors are weighted in the objective function. 
We then propose to estimate optimal weights for the CSGE objective function by comparing the variance of the linear model's predictions during daytime hours versus nighttime hours. 
The variance of the aggregate model's errors during nighttime hours reflects only load predictions; while the variance during daytime hours contains errors from both the PV and load predictions. 
Assuming that the errors from each model are independent, we use this information to estimate variances for prediction errors from the load and PV models separately. 
We then define weights for the CSGE objective function, $\alpha^*$ and $\beta^*$, to be proportional to the inverse of the expected variance of errors from each model. 
Our sensitivity results suggest that we reach an optimal disaggregation performance by using values of $\alpha^*$ and $\beta^*$ learned from the variances. 

Finally, we study the change in performance of the model when measurements are recorded at frequencies slower than the current sampling rate of once per minute. We observe that the performance of the CSGE improves drastically until 0.2 samples per minute. However, we do not see a significant improvement in the performance frequencies faster than 0.2 samples per minute. We believe that this is due to the low pass filter applied to the irradiance proxy to filter out short-term cloud cover not seen universally across the circuit. 

Although we believe that the CSGE is a generic strategy that can be applied to any distribution network given the necessary inputs, we would like to further study the impact of (i) the size of PV behind the substation, (ii) the volatility in PV generation at each location, (iii) spatial variability in PV locations, and (iv) inverter participation in reactive power control to the estimation accuracy and the validity of assumptions made in CSGE. We expect that, in some cases, the volatility can act as a disturbance to the system and improve disaggregation performance.

In addition, we would like to explore using more generally available regressors for predicting PV generation, such as clear sky irradiance and/or satellite/ground irradiance measurements, and their corresponding impact on disaggregation performance. These measurement points can be used when proxy irradiance measurements, similar to the one used in this paper, are not available.

\section*{Acknowledgement}
The authors would like to thank Riverside Public Utility for the data used in this study. This work was partially funded by U.S. Department of Energy's Advanced Research Projects Agency-Energy (ARPA-E), under Contract No. DE-AC02-05CH11231. 
\bibliography{./biblo}
\bibliographystyle{ieeetr}
\appendix
\section{Proofs}
\subsection{CSSE given parameters $\Theta$}
In this appendix we present a closed form solution recovering optimal source signals, $Y$,  when given the rest of the parameters of an $\ell_2$ CSSE model: $\Theta$. 
This proof is motivated by situations in which a CSSE model is  fit to one set of data but used to disaggregate others. 
A few examples are (A) when fitting models to only a subset of data in order to save computation time, and (B) performing online disaggregation without having to continually update all parameters. 
Equation~\ref{eq:reduced_problem_giventheta} shows the $\ell_2$ CSSE model in this special case where all $\Theta_i$ are known. $Y_i^*$ is the optimizing value of the source signal $Y_i$. 

\begin{equation}
\begin{aligned}
Y^*_i = \text{arg}\min_{Y | \Theta}& \sum_{i=1}^N \alpha_i || Y_i - X_i \Theta_i ||_2\\
 \text{ subject to, \hspace{3mm}}
& \sum_i Y_i = \bar Y
\end{aligned}
\label{eq:reduced_problem_giventheta}
\end{equation}

We find that, under these conditions, the optimal source signal is described in equation~\eqref{eqn:optimalsource}. 
Effectively, the optimal source signal is the prediction of its linear model, $X_j \Theta_j$, with the addition of some portion of the residual from the full model's prediction of $\bar Y$. 

For each $Y_i$, by taking gradients, the minimization problem presented in equation~\ref{eq:reduced_problem_giventheta} results in the following set of equations:
\begin{equation}
    \lambda=2\alpha_i(Y_i-X_i \Theta_i), \hskip10pt i \in [1,N]
\end{equation}
where $\lambda$ $\in$ $R^T$ is the vector of Lagrange multipliers. We know that the corresponding solutions must satisfy 
\begin{equation}
\sum_i^N Y_i -\bar Y = 0
\end{equation}
Hence for all signals we can write the following equation:
\begin{equation}
\frac{\lambda}{2}\sum_i^N\frac{1}{\alpha_i} + \sum_i^NX_i \Theta_i =  \bar Y
\end{equation}

Calculating $\lambda$ and substituting it in for any arbitrary reconstructed signal $Y_j$ the optimal $Y^*_j$ can be obtained as follows:

\begin{align}
\label{eqn:optimalsource}
Y_j^* &= X_j \Theta_j + \beta_j \left( \bar Y - \sum_{i=1}^N X_i \Theta_i \right)\\
\nonumber
\text{where,}\\
\label{e:errorprop}
\beta_j &= \frac{\alpha_j^{-1}}{\sum_{i=1}^N \alpha_i^{-1}}
\end{align}

Thus, the portion of the residual that is assigned to each source is dependent on the weights, $\alpha$, as shown in Eq~\eqref{e:errorprop}.

\subsection{Optimal value of $\Theta$ is independent of weights} \label{sec:alpha_beta}
In this section we show that the optimal coefficients in $\Theta$ are independent of weights, and thus can be found once and then used when determining optimal weighting. 
For this proof we break the optimization into two stages, shown in Eq~\ref{eqn:twostagemin}; where
we minimize the objective function over the source signals, $Y$ given $\Theta$ within a minimization over $\Theta$. 

\begin{align}
\label{eqn:twostagemin}
\min_\Theta \min_{Y|\Theta}  \sum_{i=1}^N  \alpha_i || Y_i -  X_i \Theta_i ||_2\\
\end{align}

Having already found closed form solutions for $Y|\Theta$ in Equation~\eqref{eqn:optimalsource}, we can substitute these solutions in for $Y_i$ to solve the inner minimization. 
Equation~\eqref{eqn:thetamin1}, shows the result of this substitution. 
In equation~\eqref{eqn:thetamin2} we change our variables to make maximizing over all parameters in $\Theta$ more straightforward; where $\bar \Theta = [\Theta_1, \Theta_2, \ldots \Theta_N]$, and $\bar X = [X_1, X_2, \ldots, X_N]$. As a result $\sum_{i=1}^N X_i \Theta_i = \bar X \bar \Theta$.

\begin{align}
\label{eqn:thetamin1}
\min_\Theta  \sum_{i=1}^N  \alpha_i || \beta_i( \bar Y - \sum_{i=1}^N X_i \Theta_i ) ||_2\\
\label{eqn:thetamin2}
\min_\Theta   \sum_{i=1}^N  \alpha_i || \beta_i( \bar Y - \bar X \bar \Theta) ||_2
\end{align}

Equations~\eqref{eqn:thetasolve1}~and~\eqref{eqn:thetasolve2} find the minimizing values of $\Theta$ using first order conditions. 
We see that all $\alpha$ values drop out and the optimal solution is equivalent to OLS regression. 
\begin{align}
0 = \sum_{i=1}^{N} \alpha_i \beta_i \bar X^T (Y - \bar X \bar \Theta)
\label{eqn:thetasolve1}\\
\label{eqn:thetasolve2}
\bar \Theta = (\bar X^T \bar X)^{-1} \bar X^T \bar Y
\end{align}
\end{document}